\documentclass[11pt]{article}

\usepackage[final]{acl}

\usepackage{times}
\usepackage{latexsym}

\usepackage[T1]{fontenc}

\usepackage[utf8]{inputenc}

\usepackage{microtype}
\usepackage{inconsolata}

\usepackage{graphicx}

\usepackage{booktabs}

\usepackage{multirow}

\usepackage{pifont}
\usepackage{xcolor}
\newcommand{\cmark}{\textcolor{green!60!black}{\ding{51}}} %
\newcommand{\xmark}{\textcolor{red}{\ding{55}}}            %

\usepackage{tikz}

\usepackage{amssymb}

\usepackage{makecell}

\usepackage{booktabs}

\newcommand*\circled[1]{\tikz[baseline=(char.base)]{
            \node[shape=circle,draw,inner sep=.6pt] (char) {#1};}}

\usepackage{xcolor}

\newcommand{\githuburl}{\url{https://github.com/jofulab/versionrag}}
\newcommand{\dataseturl}{\url{https://huggingface.co/VersionRAG}}

\usepackage{float}

\usepackage{stfloats}

\usepackage{placeins}

\newsavebox{\codebox}
\newlength{\codeinset}
\newenvironment{codeblock}
  {\par\addvspace{0.8\baselineskip}%
   \begin{lrbox}{\codebox}%
   \begin{minipage}{\dimexpr\linewidth-2\codeinset-2\fboxsep\relax}%
   \scriptsize}
  {\end{minipage}\end{lrbox}%
   \noindent\hspace*{\codeinset}\colorbox{black!5}{\usebox{\codebox}}%
   \par\addvspace{0.8\baselineskip}}

\newcommand{\rmspace}{\vspace{-0ex}}

\title{VersionRAG: Version-Aware Retrieval-Augmented Generation\\ for Evolving Documents}

\author{
  \textbf{Linus Stuhlmann\textsuperscript{1,2}}, \textbf{Kajaluxan Mathitharan\textsuperscript{1}}, \\ \textbf{Daniel Huwiler\textsuperscript{1}},
  \textbf{Kurt Stockinger\textsuperscript{1}}, \textbf{Jonathan Fürst\textsuperscript{1}}\thanks{Corresponding author: \texttt{jonathan.fuerst@zhaw.ch}}
\\
  \textsuperscript{1}Zurich University of Applied Sciences, Switzerland \\
  \textsuperscript{2}Institute of Science Tokyo, Japan\\
  \texttt{stuhlmann.l.aa@m.titech.ac.jp} \\
  \texttt{kajaluxan@mathitharan.ch, daniel.huwiler@proton.me} \\
  \texttt{kurt.stockinger@zhaw.ch, jonathan.fuerst@zhaw.ch}
}

\begin{document}
\maketitle

\begin{abstract}
Versioned corpora hold the same fact in mutually contradicting states, and the version identifier contributes only a weak signal for embedding similarity to separate them. Therefore, a Retrieval-Augmented Generation (RAG) system, when asked about a single version, usually retrieves all versions with similar scores. The result is that the answer can be arbitrary, rather than the correct one. To address this problem, we present VersionRAG, which models a text corpus as an explicit version lineage and routes queries to the relevant versions by intent. We evaluate our approach on two new datasets: VersionQA-Software (869 questions, 882 document versions) and VersionQA-Legal (547 questions, 355 versions of amended UK legislation). VersionRAG is the most accurate system in both domains across four LLMs and four baselines, reaching up to $94.4\%$ accuracy in the question-answering task. Our ablations show that the version lineage alone contributes $+21$ to $+47$ points over flat retrieval. A change layer built on it adds a further $+12$ to $+15$ points on \textit{explicit-change} and $+21$ to $+36$ on \textit{implicit-change questions}. We find that indexing the changes in the documents is critical and outperforms both generic entity graphs and temporal entity graphs.
\end{abstract}

\section{Introduction}

\texttt{torch.nn.Transformer} gained a \texttt{bias} parameter in PyTorch 2.1, defaulting to \texttt{True}, while version 2.0 has none. When asking a standard Retrieval-Augmented Generation (RAG) system \cite{fan2024surveyragmeetingllms,gao2023retrieval} about this \texttt{bias} parameter for version 2.0, the pages of 2.1 to 2.5 fill the top of the ranking, because the version identifier in the query is too weak a signal for the embedding to favor the 2.0 page (Figure~\ref{fig:version-blind}a). Versioning is ubiquitous in technical
documentation, API references, legal codes and standards. Wherever versioning occurs,
the corpus holds the \emph{same} fact in multiple, potentially mutually conflicting states,
so even a minor version change can invalidate an answer. This raises the question
underlying this paper: \emph{How should a RAG system represent and access
knowledge that exists in multiple, conflicting states due to document
versioning?} Existing methods have little mechanism for such a distinction due to the following challenges:

\begin{figure}[t]
    \centering
    \includegraphics[width=1.0\columnwidth]{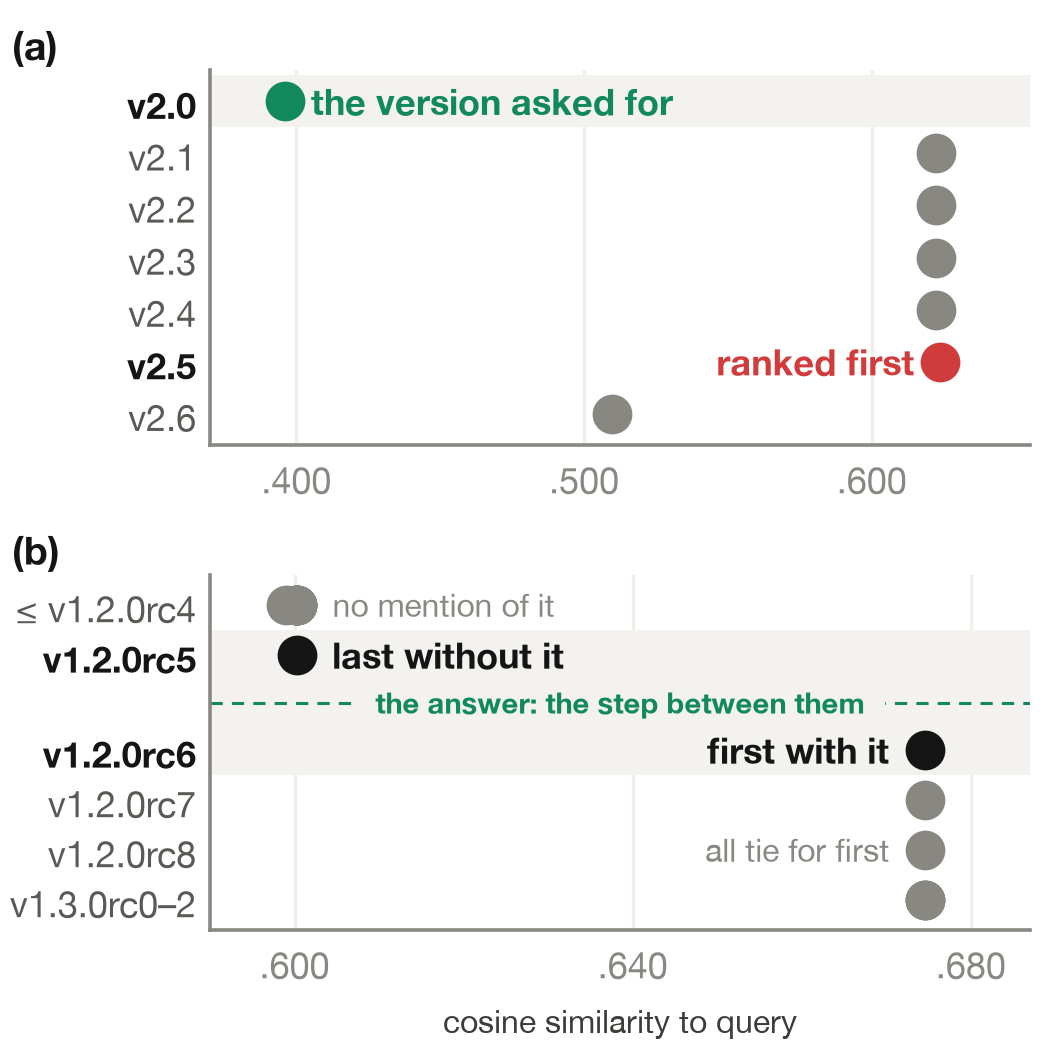}
\caption{Two failure modes of standard RAG on versioned documents.
\textbf{(a) Version conflation} for \emph{``What is the default value of the bias parameter of torch.nn.Transformer in PyTorch 2.0?''}. Every later page documents the parameter and outranks 2.0, which has none. \textbf{(b) Untracked implicit change} for \emph{``When
was NVFP4 KV Cache added to TensorRT-LLM?''}. Every version from v1.2.0rc6 onward contains the feature and all of them tie in retrieval, but none states the transition. The answer is the step from v1.2.0rc5, the last version without it, to v1.2.0rc6, which no single chunk contains.}
\label{fig:version-blind}
\rmspace
\end{figure}

\noindent \textbf{Challenge 1: Version Conflation (Figure~\ref{fig:version-blind}a).}
When asked about version 2.0, standard RAG ranks the 2.0 page last of seven: the 2.1 to 2.5 pages score highest, and their top chunk documents \texttt{bias} with default \texttt{True}, so the retrieved context describes a parameter that 2.0 does not have. \textit{Version conflation yields
incorrect or contradictory answers to version-specific queries.}

\noindent \textbf{Challenge 2: No Tracking of Implicit Changes (Figure~\ref{fig:version-blind}b).}
The challenge extends beyond version identification. When asked \textit{``When was NVFP4 KV Cache added to TensorRT-LLM?''}, a system must reason about state transitions. An \emph{implicit} change is a fact that does not exist in \emph{a single document}. Every version from v1.2.0rc6 onward documents the feature and v1.2.0rc5 does not, but no page states the transition: ``added in version $X$'' describes the boundary between two versions and appears as a statement of fact in neither. 
Graph-based retrieval systems such as GraphRAG~\cite{edge2024localglobalgraphrag} model semantic relationships between concepts but do not help, since version transitions are not entity relations extractable from any single chunk. Likewise, temporal graph extensions such as TG-RAG~\cite{han2025tgrag} use timestamp facts to disambiguate them, but version order is not time order. E.g., Python 3.8.10 and 3.9.5 shipped on the same day, and the 3.6.15 documentation postdates the 3.8.10 documentation.
\textit{Untracked implicit changes yield silently wrong answers to version-transition queries.}

To address both challenges, we present \textbf{VersionRAG (Version-Aware Retrieval-Augmented Generation for Evolving Documents)}, a retrieval-augmented generation framework that explicitly models document versioning through a structured graph representation. Unlike existing approaches that treat documents as static entities, VersionRAG constructs a hierarchical graph during indexing that captures: (1) version sequences and relationships, (2) explicit and implicit changes between versions, and (3) version-specific content boundaries. Queries are classified by intent, either (i) content retrieval, (ii) version listing, or (iii) change retrieval, and are routed through specialized retrieval paths with version-aware filtering.

We evaluate VersionRAG on two newly created datasets: \textbf{VersionQA-Software} (869 questions over 882 document versions of open-source software documentation) and \textbf{VersionQA-Legal} (547 questions over 355 versions of amended UK legislation), a domain whose versioning is structurally different, as amendments replace changelogs. Against Naive RAG, the temporal-graph method TG-RAG~\cite{han2025tgrag}, a component ablation of our own system in both domains, and against GraphRAG~\cite{neo4j_graphrag_docs} on VersionQA-Software, \textit{VersionRAG is the most accurate system in all eight domain$\times$backbone cells} (up to 94.4\% accuracy on software), with every pairwise advantage on software significant under Holm-corrected McNemar tests ($p<10^{-4}$).

Beyond the ranking, our component ablation localizes that advantage. A version-shaped index with version-scoped filtering lifts flat retrieval by 21 to 47 points. Moreover, a graph-free metadata filter over the same index reaches 85.5\% accuracy against 86.1\% for the graph ablation ($p=0.48$). Hence, what carries the gain is the version-aware \emph{representation} rather than the graph model. The advantage of the graph is the materialized and indexed change layer, which adds implicit-change awareness for facts that exist in no document. %

Overall, we make four contributions:
\circled{1} \textbf{Versioned document QA task:} We formalize the versioned document QA task and identify three fundamental query types that require distinct retrieval strategies, providing a framework for future research in this area (Section~\ref{sec:problem}).
\circled{2} \textbf{VersionRAG:} We propose VersionRAG, a retrieval framework that indexes a corpus as an explicit version lineage with materialized change facts, enabling accurate reasoning over document evolution while maintaining efficiency (Section~\ref{sec:versionrag}).
 \circled{3} \textbf{VersionQA Datasets:} We release VersionQA-Software (869 questions, 882 document versions) and VersionQA-Legal (547 questions, 355 versions), the first datasets for the task, spanning two structurally different versioning regimes (Section~\ref{sec:dataset}). 
\circled{4} \textbf{Evaluation:} VersionRAG is the most accurate system overall,
reaching 94.4\% accuracy and beating flat retrieval by 27--54 points on software (every
pairwise advantage Holm-corrected $p<10^{-4}$), at 3--4\% of GraphRAG's
index-time LLM token cost (Section~\ref{sec:results}).
All our code (\githuburl) and datasets (\dataseturl) are publicly available.

\section{Related Work}
\label{sec:related}

\noindent \textbf{Retrieval-Augmented Generation.}
RAG grounds LLMs in external knowledge \cite{lewis2021retrievalaugmentedgenerationknowledgeintensivenlp,gao2023retrieval}, and advances such as Active RAG \cite{jiang2023active} and R²AG \cite{ye2024r2ag} refine retrieval strategy but assume static corpora without version constraints. Agentic RAG and deep-search systems \cite{singh2025agenticragsurvey} improve the control loop over a fixed corpus, not version conflation: an agent explores more documents but still sees conflicting versions, so VersionRAG is orthogonal and can serve as a version-consistent back-end. Long-context LLMs face the same gap (Section~\ref{sub:problem}): implicit changes appear in no document and must be inferred from version pairs, and recent work confirms standard RAG fails on outdated information \cite{ouyang2025hohdynamicbenchmarkevaluating}.

\noindent \textbf{Graph-Enhanced Retrieval.}
GraphRAG and its variants enhance retrieval through structured knowledge representations \cite{zhang2025surveygraphretrievalaugmentedgeneration}. While GraphRAG excels at multi-hop reasoning, systematic comparisons \cite{han2025ragvsgraphragsystematic} show it struggles with temporal dynamics; in our experiments a generic entity graph does not close the gap at scale (Section~\ref{sec:results}). Efficiency-oriented variants such as LightRAG \cite{guo2024lightrag} and HippoRAG \cite{gutierrez2024hipporag} improve graph construction cost and quality, but retain the same flat entity-relation ontology: without explicit version sequences and change nodes, they face the same fundamental limitation on version-sensitive queries, and because they still extract per chunk they lower the constant of index-time cost without changing how it scales with the number of versions (Section~\ref{sub:graphfail}).

\noindent \textbf{Temporal and Version-Aware Approaches.}
Existing temporal approaches address different challenges than versioned documents. Time-Aware Language Models \cite{dhingra2022timeaware} modify LM internals to handle facts that expire, but require retraining and operate in continuous time rather than over discrete version identifiers. Temporal KGQA systems like CRONKGQA \cite{saxena2021temporal} assume pre-existing temporal knowledge graphs rather than extracting version structure from documents; adapting them to our setting would essentially require first building a version graph like ours. TempRALM \cite{gade2024itstimeincorporatingtemporality} focuses on chronological grounding (\textit{``What happened in 2023?''}) rather than version-specific retrieval (\textit{``What changed in version 2.3?''}). Closest to our setting, TG-RAG \cite{han2025tgrag} models corpora as a bi-level \emph{temporal} graph with timestamped relations and hierarchical time nodes. We include it as a baseline and find that the temporal axis does not transfer to version lineages: semantic version identifiers are not dates (Section~\ref{sub:graphfail}).

\noindent \textbf{Change Mining.}
The software engineering community has long studied fine-grained change extraction, e.g., Change Distilling \cite{fluri2007changedistilling} and RefactoringMiner \cite{tsantalis2022refactoringminer}, as well as semantic versioning conventions for tagging releases. They are complementary to VersionRAG, whose change-extraction component is modular enough to take such a miner as a back-end (Section~\ref{sec:versionrag}).

\noindent \textbf{Comparison with Existing Approaches.}
To our knowledge, VersionRAG is the first system specifically designed for retrieval-augmented generation over versioned documents; Appendix~\ref{app:related} summarizes how it combines version awareness, change tracking, and hybrid retrieval (none of which the methods above provide together, Table~\ref{tab:rag-comparison}), and discusses further related work.

\section{Problem and Data Structure}
\label{sec:problem}

We first define the problem of version-aware RAG with its three fundamental query types and then derive our data structure.

\subsection{Problem Formulation}
\label{sub:problem}
Let $\mathcal{D} = \{d_1, \dots, d_n\}$ be a collection of documents, where each document $d$ exists in a totally ordered sequence of versions $V_d = (v_1 \prec v_2 \prec \dots \prec v_{m_d})$ and $c(d,v)$ denotes the text of $d$ at version $v$. Chunking the corpus yields units $x \in \mathcal{X}$, each with provenance $\mathrm{meta}(x) = (d, v)$. For each consecutive pair $(v_j, v_{j+1})$, the \emph{change set} $\Delta(d, v_j, v_{j+1})$ is the set of atomic change facts $\delta$ (additions, removals, modifications) between the two states. A change fact is \emph{explicit} if some chunk states it (e.g., a changelog entry), and \emph{implicit} if no chunk in $\mathcal{X}$ entails it: $\delta$ describes the \emph{boundary} between $c(d,v_j)$ and $c(d,v_{j+1})$ and appears in neither text.

Given a query $q$, versioned document QA requires answers of three distinct types: (1) \textit{Content Retrieval}: information from $c(d, v)$ for a version constraint $v$ stated or implied by $q$; (2) \textit{Version Retrieval}: the version inventory $V_d$ or its metadata; and (3) \textit{Change Retrieval}: a change fact $\delta \in \Delta$. This typing makes the two failure modes above precise: $\mathrm{sim}(q, x)$ is \emph{effectively version-invariant}, so near-duplicate chunks from different versions score alike and the context mixes versions regardless of the constraint in $q$. Note that for implicit change queries \emph{no} $x \in \mathcal{X}$ entails the answer, so no retriever over $\mathcal{X}$ can supply it at any context length. Version constraints must therefore be enforced as hard filters on $\mathrm{meta}(x)$ rather than soft similarity. Moreover, implicit change facts, which no chunk entails, must be derived from version pairs rather than retrieved; VersionRAG materializes them at indexing time.

\subsection{Version-Aware Graph Structure}
\begin{figure}[t]
    \centering
    \includegraphics[width=1.0\columnwidth]{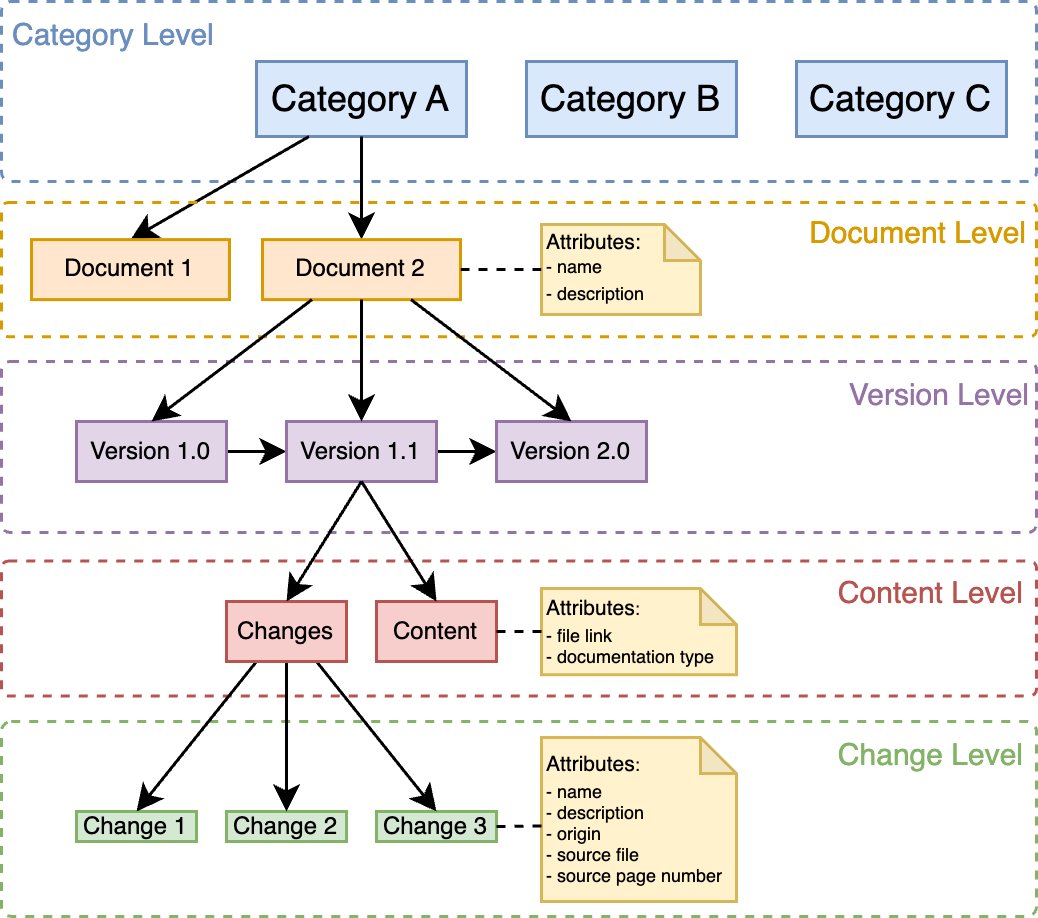}
    \caption{Version-aware graph structure with hierarchical organization from categories to individual changes, enabling precise version-specific retrieval.}
    \label{fig:versiongraph}
    \rmspace
\end{figure}

VersionRAG realizes both requirements in a typed hierarchical graph $G = (N, E)$ constructed during indexing (Figure~\ref{fig:versiongraph}). The node set is partitioned into five types, $N = N_{\mathrm{cat}} \cup N_{\mathrm{doc}} \cup N_{\mathrm{ver}} \cup N_{\mathrm{cont}} \cup N_{\mathrm{chg}}$: \textit{Category nodes} group documents semantically; \textit{Document nodes} represent each $d \in \mathcal{D}$; \textit{Version nodes} represent each $v \in V_d$; \textit{Content nodes} hold references to the embedded chunks $x$ with $\mathrm{meta}(x)$ as filterable payload; and \textit{Change nodes} materialize the change facts $\delta \in \Delta(d, v_j, v_{j+1})$, extracted from changelogs (explicit) or computed by diffing consecutive versions (implicit). Edges are typed accordingly: containment edges follow the hierarchy ($N_{\mathrm{cat}} \to N_{\mathrm{doc}} \to N_{\mathrm{ver}} \to N_{\mathrm{cont}}$), \textsc{next} edges encode the order $\prec$ over each $V_d$, and each Change node connects the version pair $(v_j, v_{j+1})$ it describes. Each query type thus maps to one access path: (i) a version-constrained content query resolves the Version node and runs a vector search filtered on $\mathrm{meta}(x)$; (ii) a version-listing query enumerates the document's Version nodes; (iii) a change query looks up its Change nodes, where implicit changes are stored as nodes of their own and can be retrieved directly.

\section{VersionRAG}
\label{sec:versionrag}

\begin{figure*}[ht]
   \centering
   \includegraphics[width=0.9\linewidth]{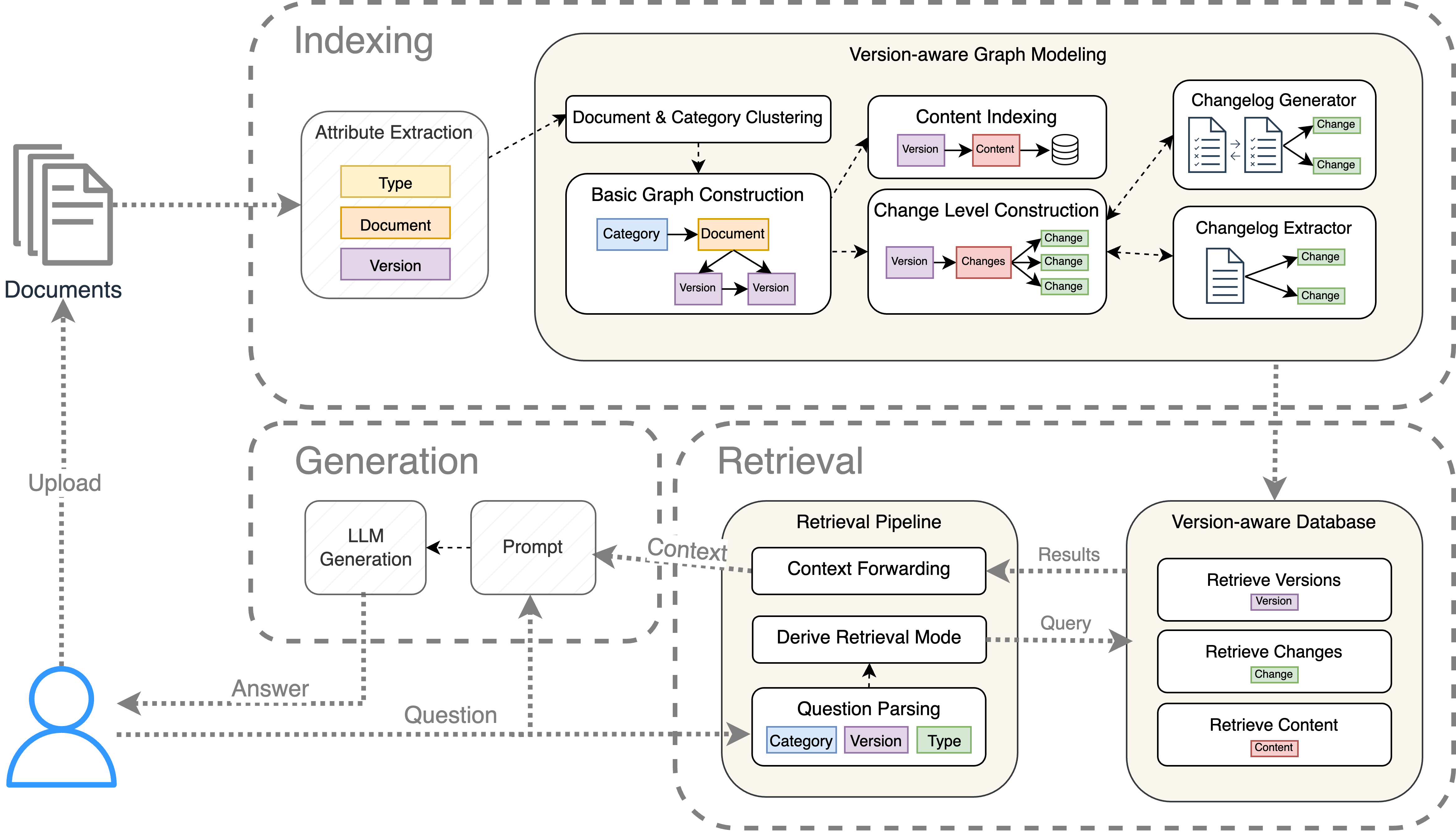}
   \caption{VersionRAG framework with three main components: indexing constructs the version-aware graph, retrieval routes queries through appropriate paths, and generation produces answers using retrieved context.}
   \label{fig:framework}
       \rmspace
\end{figure*}

VersionRAG has three parts: \textbf{indexing}, which builds the version-aware graph; \textbf{retrieval}, which detects intent, selects a retrieval mode and applies version-aware filtering; and \textbf{generation} over version-specific context (Figure~\ref{fig:framework}).

\subsection{Indexing Pipeline}
Indexing transforms raw documents into the version-aware graph through five steps. First, \textit{attribute extraction} obtains document metadata (title, version, type). If documents carry machine-readable version markers (version-bearing filenames or explicit \texttt{Version:} headers), metadata is obtained by \emph{parsing} at zero LLM cost, with per-project normalization of the raw version suffix. Otherwise (PDF documentation, layouts that do not encode versions, versions stated only in prose), an LLM extracts the same metadata. We ran both approaches over the software corpus and find them equivalent within $\pm 3$\,pp (Appendix~\ref{app:attrpath}), so our results do not depend on which approach provides the metadata.

Second, \textit{document clustering} groups versions of the same document and organizes them into categories using semantic similarity. Third, \textit{graph construction} creates the hierarchical structure with proper version sequencing. Fourth, \textit{content indexing} chunks documents and stores embeddings in a vector database with version metadata. Last, \textit{change extraction} identifies modifications either from explicit changelogs or by computing differences between consecutive versions using DeepDiff~\cite{Dehpour_DeepDiff}, and then generates semantic descriptions of changes with an LLM. Each LLM-based component is modular and local to its subtask: rule-based extractors or domain-specific change miners~\cite{fluri2007changedistilling,tsantalis2022refactoringminer} can be substituted without changing the graph schema or retrieval logic.

\subsection{Hybrid Retrieval Strategy}
VersionRAG employs a query-aware retrieval strategy that routes queries through appropriate retrieval paths based on their intent. The retrieval pipeline consists of three components: (1) \textit{Query parsing} uses an LLM to classify the query type and to extract relevant parameters (category, document, version). (2) \textit{Retrieval mode selection} determines whether to use graph traversal (for version/change queries) or vector search (for content queries). (3) \textit{Version-aware filtering} constrains vector search to specific versions when needed, ensuring temporal accuracy. For example, \textit{``What is the stability of assert.CallTracker in Node.js 15.14.0?''} is parsed as version-specific content retrieval with parameters (\texttt{doc=assert}, \texttt{version=15.14.0}); the router resolves the Version node and issues a vector search whose payload filter admits only chunks of that version. On the other hand, the query \textit{``When was assert.CallTracker removed?''} is routed to graph traversal over Change nodes (see \ref{app:worked} for an example per query type).

\subsection{Generation}
After retrieval, the generation component synthesizes answers using the retrieved context. The LLM receives both the original query and version-specific context, with explicit instructions to ground responses in the provided information. Unlike other RAGs where context may contain version inconsistent information, VersionRAG ensures that the retrieved context aligns with the query's version requirements, enabling accurate and consistent answer generation.

\section{Dataset Construction}
\label{sec:dataset}

We create two version-aware datasets. \textbf{VersionQA-Software} comprises 869 questions over 882 document versions,
covering 50 documentation topics from open-source projects, of which 313 files are versioned
documentation and 569 are changelogs, drawn from disjoint project sets so that explicit and
implicit change items have disjoint sources. It carries the full evaluation: every system on
every backbone. \textbf{VersionQA-Legal} comprises 547 questions over 355
point-in-time versions of five amended UK Public General Acts \cite{uk_legislation}, where
amendments arrive as in-place edits to dense, self-similar statutory prose rather than as release
notes; we use it to ask whether version-shaped retrieval transfers to that regime at all, and
read its absolute accuracies as a harder operating point. Both datasets use the same six categories, which refine the three query types of
Section~\ref{sub:problem} and are sampled near-uniformly by design, to measure each capability rather than to
mimic real query traffic. Four require version-aware reasoning; sourcing, version selection,
per-category counts (Table~\ref{tab:querycategories}), category definitions, and examples are in
Appendix~\ref{app:datasets}.

We generate questions with Claude Opus 4.8 per category from sampled sources using author-written per-category templates (reproduced in Appendix~\ref{app:genprompts}) and no item is accepted on a
single generation pass: every pair is re-scored against the sources the generator saw, and items
in the three structure-dependent categories are re-validated against the index, which drops
ambiguous items and corrects wrong golds. Implicit-change items, the category our central claim
rests on, are additionally checked by hand.
These checks describe how the items were produced, not how correct the released data is. To measure that, two annotators audited a stratified
sample of 120 released golds (10 per category per domain) against the corpus: We identified two version-ordering errors in VersionQA-Software, an error rate of 1.7\% over the 120 audited items (3.3\% on software). Two samples were identified as partly correct. On the 40 items labeled by both annotators, verdicts agree on 39
(Gwet's AC1 $=0.97$). The audit also finds that 9 of 10 sampled legal implicit-change items are
stated in an earlier version's amendment annotation. Hence, in the legal set ``implicit'' holds for the
version diff but not always for the indexed text (Appendix~\ref{app:goldaudit}).

\section{Experimental Evaluation}
\label{sec:results}

Our analysis structures around the central question: \emph{Where does version-aware accuracy come from?}

\subsection{Experimental Setup}

\noindent \textbf{Systems.}
(1) \textit{Naive RAG}
\cite{gao2023retrieval}: 512-token chunks, dense retrieval, no version awareness.
(2) \textit{GraphRAG} \cite{neo4j_graphrag_docs}: Neo4j's GraphRAG package (\texttt{neo4j-graphrag})
(\texttt{SimpleKGPipeline} with \texttt{HybridCypherRetriever}), a generic entity--relation graph
built by per-chunk LLM extraction, which tests whether general graph structure helps.
(3) \textit{TG-RAG} \cite{han2025tgrag}: a bi-level temporal hierarchy over timestamped quadruples.
(4) \textit{Metadata-Filter}: a graph-free version-aware baseline that resolves the requested document
and version with the same query parser, then reads catalog and version listings from vector-store
payloads and retrieves densely within the resolved version, with the graph database stopped; it
isolates version awareness from graph structure.
(5) \textit{VersionRAG w/o Change}: identical to VersionRAG except that no
Change nodes are built; it retains the document--version hierarchy, its ordering, and
version-scoped vector search.
(6) \textit{VersionRAG}: the full system.

\noindent \textbf{Configuration.}
We evaluate four generator backbones (\texttt{gpt-5.4-nano} \cite{openai2026gpt54nano},
\texttt{gpt-5.4-mini} \cite{openai2026gpt54mini}, \texttt{mistral-small-2603} (Mistral Small 4, 119B MoE)
\cite{mistralsmall4} and \texttt{qwen3-32b} \cite{yang2025qwen3}) with
\texttt{text-embedding-3-small} \cite{openai2024embedding}. For all experiments one model
serves both indexing and generation. VersionRAG uses Neo4j \cite{neo4j2025} and Qdrant
\cite{qdrant2025}, line-based chunks of at most 1,024 characters without overlap, and DeepDiff
\cite{Dehpour_DeepDiff} for implicit change detection; all prompting is few-shot
(Appendix~\ref{app:implementation_details}).
We use VersionQA-Software to evaluate all systems against VersionRAG. Due to token costs, we use VersionQA-Legal only to show the generalizability of our method to new domains.

\noindent \textbf{Metrics and Statistical Protocol.}
Accuracy is the proportion of correctly answered questions, each scored by \texttt{gpt-5.4}
\cite{openai2026gpt54} as a binary judge \cite{gu2025surveyllmasajudge} restricted to
answer--gold matching. We report 95\% Wilson intervals \cite{wilson1927probable} (corroborated by
10k-resample bootstrap) and test pairwise differences with paired McNemar tests
\cite{mcnemar1947note}, Holm-corrected \cite{holm1979simple} within each domain $\times$ generator
family. To validate the judge, we re-score all 26,022 scored answers with an independent NLI
mutual-entailment metric \cite{laurer2024nli} and hand-adjudicate a blind stratified sample of
306 software answers (Section~\ref{sub:judge}).

\subsection{Overall Results}
\label{sub:mainresults}

Table~\ref{tab:mainresults} reports overall accuracy on VersionQA-Software with 95\% Wilson intervals. VersionRAG is the most accurate system for every one of the four backbones, reaching an accuracy of 94.4\% with \texttt{gpt-5.4-mini}. VersionRAG  outperforms VersionRAG w/o Change, which in turn outperforms every baseline on all four backbones with non-overlapping intervals. Hence the effect is not a single-model artifact; only the order among the three baselines varies by backbone.

\begin{table*}[t]
\centering
\small
\caption{Overall accuracy [\%] on VersionQA-Software with 95\% Wilson confidence intervals.
GraphRAG and TG-RAG run with their standard indexing and retrieval pipelines, untuned for our corpus, and are included to characterize
how non-version ontologies behave on a versioned corpus, not as tuned competitors; the weight of
our claim rests on the ablation in the last two rows (Section~\ref{sub:graphfail}).}
\begin{tabular}{lcccc}
\hline
System & \texttt{gpt-5.4-nano} & \texttt{gpt-5.4-mini} & Mistral Small 4 & \texttt{qwen3-32b} \\ \hline
Naive RAG & 44.8 [41.5--48.2] & 40.5 [37.2--43.8] & 34.3 [31.2--37.5] & 57.6 [54.3--60.9] \\
GraphRAG & 31.2 [28.2--34.4] & 24.1 [21.3--27.0] & 26.1 [23.2--29.1] & 39.0 [35.7--42.3] \\
TG-RAG & 47.1 [43.7--50.4] & 52.1 [48.8--55.5] & 40.7 [37.5--44.1] & 34.0 [30.9--37.3] \\
VersionRAG w/o Change (ours) & 82.0 [79.3--84.5] & 87.4 [85.0--89.5] & 79.3 [76.5--81.9] & 78.8 [75.9--81.4] \\
\textbf{VersionRAG (ours)} & \textbf{89.4 [87.2--91.3]} & \textbf{94.4 [92.6--95.7]} & \textbf{87.2 [84.8--89.3]} & \textbf{84.2 [81.6--86.5]} \\
\hline
\end{tabular}
\label{tab:mainresults}
\rmspace
\end{table*}

\noindent \textbf{Statistical Significance.}
Paired McNemar tests on per-question correctness and Holm-corrected within each generator family confirm the results. Quoting the \emph{worst} of the four backbone cells in each case, VersionRAG beats Naive RAG at $p=3.3\times10^{-35}$, GraphRAG at $p=1.0\times10^{-75}$, TG-RAG at $p=6.9\times10^{-96}$, and its own ablation (VersionRAG w/o Change) at $p=5.9\times10^{-5}$. The last is by far the tightest comparison, only $+5.4$ to $+7.9$ points, yet all four cells show a clear $p<10^{-4}$.

\subsection{Decomposing the Advantage: Structure vs.\ Change Layer}
\label{sub:ablation}

The ablation splits the Naive$\to$VersionRAG gap into two steps with one variable each, read off Table~\ref{tab:mainresults}.

\paragraph{Step 1: Version-lineage structure.}
Moving from flat retrieval to a version-shaped graph with version-scoped vector filtering yields an improvement of 21.1 to 46.9 points, while adding the LLM change layer on top contributes a further +5.4 to +7.9. Most of VersionRAG's advantage is therefore carried by the \emph{version-lineage representation}. The ablation (VersionRAG w/o Change) is the second-best system we test, far above a generic entity graph, which does not substitute for version shape (Section~\ref{sub:graphfail}). What carries the gain is not an elaborate search procedure but an index that satisfies three requirements simultaneously: per-chunk $(d,v)$ provenance, a router that resolves a question to a (document, version) pair, and a canonically ordered lineage over which version sets can be enumerated and relative references resolved. Naive RAG has none of the three, and no amount of retrieval tuning supplies them, because they are properties of the representation rather than of the search over it.

\paragraph{The structure, not the graph.}
To separate version awareness from graph machinery, we add a graph-free \emph{Metadata-Filter}
arm that resolves the query to a (document, version) pair and filters the vector store on chunk
payloads, with the graph database stopped. On one shared index with \texttt{gpt-5.4-mini} it
reaches an accuracy of 85.5\%, against 86.1\% for VersionRAG w/o Change re-run on that same index
(87.4\% in Table~\ref{tab:mainresults}): a gap of 0.6 points that is not
significant ($p=0.48$), while the full system reaches 94.4\%. The version-aware
\emph{representation}, not the graph, therefore carries the gain over flat retrieval on content
and version-listing queries. The \textit{change layer} yields the main advantage of the graph, worth $+14$ points on
explicit and $+27$ on implicit change in the same comparison. We ran this arm on
\texttt{gpt-5.4-mini} only, so it establishes the decomposition for that cell rather than across
backbones.

\begin{table}[H]
\centering
\small
\caption{Capability-level accuracy [\%] on VersionQA-Software, VersionRAG $\to$ VersionRAG w/o Change. The loss concentrates in \emph{implicit} change.}
\resizebox{\columnwidth}{!}{%
\begin{tabular}{lcccc}
\hline
Generator & Content & \makecell{Version\\Listing} & \makecell{Change\\explicit} & \makecell{\textbf{Change}\\\textbf{implicit}} \\ \hline
\texttt{gpt-5.4-nano} & 89 $\to$ 89 & 98 $\to$ 98 & 95 $\to$ 80 & \textbf{77 $\to$ 48} \\
\texttt{gpt-5.4-mini} & 95 $\to$ 95 & 99 $\to$ 99 & 97 $\to$ 83 & \textbf{86 $\to$ 59} \\
Mistral Small 4 & 88 $\to$ 88 & 98 $\to$ 98 & 90 $\to$ 78 & \textbf{73 $\to$ 37} \\
\texttt{qwen3-32b} & 84 $\to$ 85 & 93 $\to$ 92 & 85 $\to$ 72 & \textbf{76 $\to$ 55} \\ \hline
mean $\Delta$ & --- & --- & $-13.6$ & $\mathbf{-28.3}$ \\
\hline
\end{tabular}}
\label{tab:capability}
\rmspace
\end{table}

\paragraph{Step 2: Change layer.}
Table~\ref{tab:capability} shows the results of VersionRAG vs. VersionRAG w/o the change layer evaluated on the software dataset. Removing the change layer changes content and version-listing accuracy by at most 1 point on every generator, but lowers explicit and implicit change accuracy by 13.6 and 28.3 points on average.

\subsection{Why Implicit Change Is Materialized at Index Time}
\label{sub:mechanism}

The concentration of the ablation gap in \emph{implicit} change has a precise mechanism. An implicit change is a fact that exists in no document: ``\texttt{PYTHON\_PRESITE} was added in 3.13'' describes the \emph{delta} between 3.12 and 3.13 and appears in neither version's text. The ablation's generator is not version-blind, since every retrieved chunk is labeled with its document and version, but presence is not first-appearance: a chunk tagged 3.13 that contains the flag proves presence \emph{in} 3.13, not absence in 3.12. A single chunk is a point; ``added in $X$'' is a boundary between two points. Because top-$k$ retrieval mixes near-identical sections across versions, the model \emph{guesses} the boundary: $\sim$88\% of the ablation's software change failures are confident-but-wrong answers rather than abstentions (in one observed failure the model confidently answered 3.7; gold: 3.13). VersionRAG's Change nodes pre-compute the boundary once at index time rather than per query, and the router maps change queries directly to them.

Paired question-level evidence supports this reading: across generators the change layer \emph{fixes} 3--12$\times$ more software change questions than it breaks, and the asymmetry is sharpest on implicit change, 151 fixes against 23 regressions in aggregate ($6.6\times$; per-backbone counts in Appendix~\ref{app:fixbreak}). Explicit changes degrade less without the layer because changelog text remains in the vector index and partially recovers the answer; implicit changes have no textual fallback: the index-time diff is the only place the fact is ever materialized.

\subsection{Where Routing Fails}
\label{sub:routing}

Query classification is shared by both version-aware arms. Scored
against the gold category of all 869 software questions, the parser has one systematic failure
(Table~\ref{tab:parseraccuracy}): explicit-change routing reaches 70.5--77.9\% accuracy, while every other
category stays at or above 96.5\% on every backbone. Explicit-change questions sit on a changelog
page and usually name the version they ask about, which also satisfies the version-specific rule.
 Implicit change is never misrouted. Per-backbone
confusion matrices are in Appendix~\ref{app:parser}.

\begin{table}[H]
\centering\small
\caption{Query-parser routing accuracy [\%] on VersionQA-Software by gold category
(CR = content retrieval). Error is concentrated almost entirely in explicit change retrieval;
every other category is at or above 96.5\% accuracy for every backbone.}
\resizebox{\columnwidth}{!}{%
\begin{tabular}{lrrrrr}
\toprule
Gold category & $n$ & nano & mini & Mistral S4 & qwen3-32b \\
\midrule
CR                 & 150 & 96.7 & 100.0 & 98.7 & 99.3 \\
CR complex         & 139 & 98.6 & 100.0 & 100.0 & 100.0 \\
CR version-specific & 143 & 97.9 & 96.5 & 99.3 & 100.0 \\
Version listing    & 145 & 99.3 & 100.0 & 100.0 & 100.0 \\
Change explicit    & 149 & 77.9 & 70.5 & 71.1 & 73.2 \\
Change implicit    & 143 & 100.0 & 100.0 & 100.0 & 100.0 \\
\midrule
Overall            & 869 & 94.9 & 94.4 & 94.7 & \textbf{95.3} \\
\bottomrule
\end{tabular}}
\label{tab:parseraccuracy}
\rmspace
\end{table}

\subsection{Robustness to Incomplete Version Histories}
\label{sub:coverage}

Real archives often lack intermediate releases. To measure what this costs, we drop intermediate versions in release order, rebuild the change layer over the remaining versions, and re-run 107 of the 143 implicit-change questions with \texttt{gpt-5.4-mini}. 

We measure scores in two different ways. \emph{Strict} accuracy requires naming
the gold version, while \emph{interval-correct} accuracy requires naming the correct boundary
between the versions that remain indexed. This is the most precise true answer available once
the gold version itself has been dropped.

\begin{table}[H]
\centering
\small
\caption{Implicit-change accuracy [\%] under thinned version histories
(\texttt{gpt-5.4-mini}, 107 questions). Strict scoring requires the gold version; interval-correct
scoring requires the correct boundary between the versions still indexed.}
\begin{tabular}{lcc}
\hline
Version coverage & Strict & Interval-correct \\ \hline
100\% & 82.2 & 82.2 \\
66\%  & 61.7 & 80.4 \\
26\%  & 19.6 & 85.0 \\
\hline
\end{tabular}
\label{tab:coverage}
\rmspace
\end{table}

Strict accuracy falls from 82.2\% to 19.6\% as version coverage drops to 26\%, while interval-correct
accuracy holds between 80.4\% and 85.0\% (Table~\ref{tab:coverage}). The strict collapse is an
artifact of the gold version no longer being in the corpus rather than a failure to localize the
change: restricted to the questions whose gold version is still indexed at 66\% coverage,
accuracy is unchanged at 84.0\%, against 84.0\% at full coverage. Missing intermediate versions therefore
cost \emph{resolution}, not correctness; the system reports the tightest boundary the surviving
corpus supports.

\subsection{Validating the LLM-as-a-Judge}
\label{sub:judge}

To rule out errors introduced by LLM-as-a-judge,  we re-scored all 26,022 scored answers with an independent NLI mutual-entailment metric (\texttt{dleemiller/ModernCE-large-nli}) \cite{moderncenli2025}. This metric does not  share any of the judge's biases. Applying the re-scoring \textbf{preserves the ranking of the evaluated systems}: VersionRAG and VersionRAG w/o Change remain the top-performing systems, while the graph baselines remain at the bottom. \textbf{The disagreement is asymmetric}: judge-correct/NLI-incorrect occurs 4,521 times against 405 for the reverse on the 17,377 software answers ($\kappa=0.457$), as expected, since NLI is the stricter surface-form test and the top systems answer more paraphrastically. Stricter is not more accurate, so we settle ground truth by hand.

We adjudicated 306 software answers against independently verified references, blind to verdict and system, stratified by judge verdict $\times$ NLI agreement so that population-weighted rates estimate judge error over all 17,377 scored software answers, not only the disagreements (Appendix~\ref{app:judgeaudit}). The judge is wrong on 2.50\% [0.22, 4.78] of the answers it accepts and 5.52\% [0.75, 10.29] of those it rejects: it does not invent correctness, and the two errors partly offset in aggregate. The errors also do not favor our systems: none of the 144 adjudicated acceptances for VersionRAG or VersionRAG w/o Change was wrong, against 9 of 30 for TG-RAG\@. %

\subsection{Why the Graph Baselines Fail}
\label{sub:graphfail}

TG-RAG has the right idea on the wrong axis: version order is not time order, so a wall-clock timeline cannot represent a version lineage, no matter how its timestamps are obtained. Its graph over our corpus holds 732 date nodes to 5 version nodes, because its normalizer rejects semantic-version strings as dates. GraphRAG fails for the neighboring reason: entity--relation triples per chunk, no version ordering, no change nodes, so at scale the neighborhood around an entity mixes \emph{all} versions and retrieval returns version-conflated context. Both are faithful implementations of methods built for a different ontology, and neither is repairable by retuning: supplying the version order, content boundaries and materialized change facts they lack would amount to rebuilding the representation we propose. The weight of our claim therefore rests on the ablation rather than on any baseline's configuration. 
\textbf{Indexing Cost.} Consecutive versions of a document share most of their text, so per-chunk extraction re-processes nearly the same content for every version: its index-time cost grows with each version, although a new version adds little information. Each additional version costs GraphRAG $135$--$175$\,K LLM tokens and TG-RAG $49$--$63$\,K, but VersionRAG only $5.2$--$5.7$\,K, spent on one attribute call per version and one change call per consecutive version pair. Over the whole corpus, GraphRAG spends $22$--$29\times$ and TG-RAG $8$--$10\times$ the $5.29$\,M tokens of corpus text, VersionRAG only $0.87$--$0.95\times$: it indexes a versioned corpus with fewer LLM tokens than the corpus contains. This follows from the version design, not from tuning: both baselines run with their standard pipelines, untuned for our corpus (Appendix~\ref{app:resources}).

\subsection{Domain Transfer: Legal}
\label{sub:legal}

On legal, flat retrieval answers one question in four, while VersionRAG answers three in four. In contrast to VersionQA-Software, VersionQA-Legal contains no changelogs at all, and amendments arrive as micro-edits buried in documents orders of magnitude longer than the edit itself, yet the version axis still triples accuracy (Table~\ref{tab:legal}). VersionRAG is the top system on all four backbones, ahead of flat retrieval by 43.1 to 53.7 points (all $p<10^{-51}$) and of TG-RAG by 42.6 to 54.4, although legal versions are ISO dates that its temporal graph accepts. The change layer's benefit is significant for Mistral Small 4 ($+8.6$ pp, $p=4\times10^{-6}$), \texttt{qwen3-32b} ($+7.5$ pp, $p=4\times10^{-5}$) and \texttt{gpt-5.4-mini} ($+4.4$ pp, $p=0.023$), and neutral for \texttt{gpt-5.4-nano} ($+1.8$ pp, n.s.).
\begin{table}[H]
\centering
\small
\caption{Overall accuracy [\%] on VersionQA-Legal. Same four backbones as
Table~\ref{tab:mainresults}; GraphRAG is characterized on software only.}
\resizebox{\columnwidth}{!}{%
\begin{tabular}{lcccc}
\hline
System & nano & mini & Mistral S4 & qwen3-32b \\ \hline
Naive RAG & 24.5 & 24.5 & 20.7 & 26.0 \\
TG-RAG & 13.2 & 32.2 & 22.9 & 18.3 \\
VersionRAG w/o Change & 65.8 & 70.4 & 65.8 & 64.5 \\
\textbf{VersionRAG} & \textbf{67.6} & \textbf{74.8} & \textbf{74.4} & \textbf{72.0} \\
\hline
\end{tabular}
}
\label{tab:legal}
\rmspace
\end{table}
\FloatBarrier

Two factors set the change layer's value: domain structure \emph{and} generator capability. Ablation-only legal accuracy is largely generator-insensitive (64.5--70.4\%), because the vector fallback feeds every generator the same insufficient context, whereas the full system's legal change accuracy orders cleanly with capability (47\% nano, 61\% mini, 65\% \texttt{qwen3-32b}, 70\% Mistral Small 4). 

Legal Change nodes are long statutory-amendment records: weaker generators cannot use them even when the right one is retrieved, which is why the change layer barely helps nano on legal. Legal remains the harder operating point for every system, since statutory prose is dense and self-similar (Appendix~\ref{app:legalhard} measures why), but the transfer holds, and the change layer improves on the ablation in three of four configurations there.

\section{Conclusion}
We presented VersionRAG, a retrieval-augmented generation framework for versioned documents, together with VersionQA-Software and VersionQA-Legal, which establish versioned document QA as a measurable task across two structurally different versioning regimes. VersionRAG is the most accurate system in every combination of domain and backbone, and the ranking holds under an independently validated judge.
Our component ablation decomposes that advantage: indexing the same corpus on the version axis adds 21 to 47 points over flat retrieval, far more than a temporal or generic entity graph, and the change layer on top adds change retrieval, concentrated in implicit changes (Section~\ref{sub:ablation}).

\section*{Acknowledgments}
This work was partially supported by DataGEMS, funded by the European Union’s Horizon Europe Research and Innovation Programme, under grant agreement number 101188416.
\clearpage

\section*{Limitations}

\paragraph{Branching Version Histories.}
Our formalization assumes a total order $v_1 < v_2 < \dots < v_m$ per document, whereas real
software versioning is frequently a DAG: release lines are maintained in parallel, and security
fixes are backported to older branches after newer ones have shipped. Our own software corpus
projects such a history onto a single order, since the Python 3.8--3.15 documentation lines
were maintained in parallel over the period the corpus covers.
The data model absorbs the generalization, because Change nodes are defined over typed edges and a
\textsc{next} edge becomes a parent-to-child relation admitting several successors; the query
semantics do not. ``The previous version'' stops being unique, and ``first added in $X$'' requires
a per-branch answer rather than a single boundary. We do not evaluate on branching histories, and
the accuracies reported here should not be read as evidence about them.

\paragraph{A Single Constraint Axis.}
We model version as the only axis along which a document varies, while real documentation often
varies along several at once (version $\times$ platform $\times$ locale $\times$ edition).
Filtering generalizes cheaply: an additional axis is one more conjunct on the chunk tags and
leaves the retrieval path unchanged. Change materialization does not, because the number of
adjacent pairs to diff grows with the product of the axes, and our change layer models no
cross-axis difference at all, so a question such as what differs between the Linux and the Windows
build of one release is outside the index we construct.

\paragraph{Fixed Embedding Model and Judge.}
All experimental cells use \texttt{text-embedding-3-small} for retrieval and \texttt{gpt-5.4} as judge; alternative embedding models and judges are not ablated. We validate the judge with an independent NLI-based metric and a blind 306-answer human audit (Section~\ref{sub:judge}), which together preserve the system ranking.

\paragraph{Gold-Answer Error Rate.}
Our gold answers are model-generated and validated against the corpus rather than authored by
domain experts. A two-annotator audit of a stratified 120-item sample puts the error rate at
1.7\% [0.5--5.9], with a further 1.7\% of golds partly correct (Appendix~\ref{app:goldaudit}),
which bounds every accuracy we report: differences narrower than that margin should not be read
as established. Agreement is measured on the 40 items both annotators labeled, which cover the two
categories that turn on version order, not on a random subset. Reported accuracies use the
original golds.

\paragraph{Change Detection.}
The DeepDiff-based comparison may miss semantic changes that do not manifest as clear textual differences (e.g., a security implication of a changed default value); in the legal domain, where every change must be recovered from a long-text diff, micro-amendments are correspondingly harder to detect.

\paragraph{Scope of the Legal Evaluation.}
VersionQA-Legal establishes that version-shaped retrieval carries over to a structurally different
versioning regime, and that is the claim we draw from it. Its questions and gold answers come from
the same automated pipeline as the software benchmark and were not reviewed by legal
practitioners. A thorough treatment of statutory QA, covering practitioner-validated items, in-force-date
semantics, and the distinction between enacted text and editorial annotation, is future work.

\paragraph{Temporal Scope and Error Propagation.}
VersionRAG handles discrete version identifiers, not continuous timestamps: questions like ``What features were added in Q3 2023?'' require mapping dates to version releases, which the current system does not support. And because the architecture is a pipeline, change-extraction errors persist in the graph until reindexing rather than being corrected at query time.

\section*{Ethical Considerations}

\paragraph{Environmental Impact.}
Indexing calls an LLM once per document version and once per consecutive version pair, totaling 0.87--0.95$\times$ the corpus size in tokens, 3--4\% of GraphRAG's index-time budget (Appendix~\ref{app:resources}).

\paragraph{Fairness and Bias.}
VersionRAG's performance depends on the underlying LLM's capabilities, inheriting any biases present in these models. Our evaluation covers English technical and legal documentation; performance may vary for other languages or domains. The change layer's benefit scales with generator capability (Section~\ref{sec:results}), which could create disparities in service quality based on available computational resources.

\paragraph{Data Privacy.}
Versioned documents often contain sensitive information about deprecated features, security vulnerabilities, or internal changes. Because our implementation sends document content to hosted LLM APIs for both indexing and generation, organizations must carefully consider which documents to include and ensure appropriate access controls on the resulting knowledge graph; a self-hosted backbone removes the first concern but not the second.

\paragraph{Transparency and Accountability.}
Residual errors, particularly in change detection, could lead to incorrect technical or legal decisions if users over-rely on system outputs. We recommend showing the source document and version of every retrieved passage, which the graph's provenance makes possible. The version-aware graph structure enhances accountability by maintaining clear provenance for all information.

\paragraph{Broader Impacts.}
Better access to versioned documentation helps developers working with evolving systems, but the same capability applies to security documentation, so deployments should keep appropriate access controls.

\paragraph{Human Annotation.}
The gold-answer and judge audits were carried out by the authors.

\paragraph{AI Assistants in Research or Writing} We have used Claude Opus models for coding, dataset construction (see Section~\ref{sec:dataset}), and refining our writing. All final text in the paper has been verified by the authors.

\bibliography{custom}

\appendix

\section{Additional Related Work}
\label{app:related}

This section extends the concise discussion in Section~\ref{sec:related}.

\subsection{Evolution of RAG Systems}

The development of retrieval-augmented generation has seen significant advances beyond the foundational work. REALM \cite{guu2020retrieval} pioneered retrieval-enhanced pre-training, while dense retrieval methods \cite{karpukhin2020dense} and efficient architectures like ColBERT \cite{khattab2020colbert} improved retrieval quality.

Recent innovations have introduced more sophisticated strategies. RICHES \cite{jain2024ragriches} unifies retrieval with generation by directly decoding document contents, eliminating the need for separate retriever and generator components. Multi-Level Information RAG \cite{adjali2024multilevel} enhances answer generation through entity retrieval and query expansion with joint-training loss. These advances demonstrate the field's progression toward more integrated and efficient architectures, though none address version-specific challenges.

\subsection{Graph-Based Retrieval Systems}

Beyond GraphRAG, several systems have explored graph-enhanced retrieval. TRACE \cite{fang2024trace} employs a KG Generator to create knowledge graphs from retrieved documents and constructs reasoning chains for multi-hop QA. Knowledge Graph-Enhanced LLMs \cite{li2024kgframework} use question decomposition and atomic retrieval to improve reasoning over structured knowledge. Q-KGR \cite{zhang2024qkgr} introduces question-guided re-scoring to eliminate noisy pathways in retrieved subgraphs.

While these approaches demonstrate the value of graph-based reasoning, they focus on semantic relationships rather than temporal or version-specific connections, and they retain the flat entity--relation ontology whose limitations on versioned corpora we quantify in Section~\ref{sec:results}. Their construction cost is also substantial, since entity extraction runs an LLM over every chunk of the corpus.

\subsection{Temporal Reasoning in NLP}

Temporal reasoning has been extensively studied across different aspects. TimeR$^4$ \cite{qian2024timer4} integrates temporal knowledge from Temporal Knowledge Graphs into LLMs through a Time-aware Retrieve-Rewrite-Retrieve-Rerank framework. Time Matters \cite{barik2024timematters} addresses temporal claim verification by extracting temporal cues and applying temporal reasoning for fact-checking. Natural Evolution-based approaches \cite{chen2024natural} model asynchronous characteristics of event evolution in temporal KGs using dual-level aggregation.

These approaches focus on continuous temporal dimensions (dates, time periods) rather than discrete version identifiers. The distinction is crucial: version transitions in technical documentation follow semantic versioning conventions (major.minor.patch) that encode compatibility and change severity, which calendar-time reasoning systems cannot capture, consistent with our TG-RAG results in Section~\ref{sec:results}.

\subsection{Domain Adaptation in RAG}

Several recent works have explored adapting RAG to specific domains. RAG-Studio \cite{mao2024ragstudio} generates domain-specific training data through self-alignment, fine-tuning both LLMs and retrievers to work cohesively for domain-specific tasks. LongRAG \cite{zhao2024longrag} introduces a dual-perspective paradigm for long-context QA, processing information at both global and local levels. Unified Active Retrieval \cite{cheng2024unified} addresses the challenge of when to retrieve through multiple orthogonal criteria.

While these systems improve domain-specific performance, they do not address the core challenge of document versioning, where the same query may have different answers depending on the version context.

\subsection{RAG Evaluation and Analysis}

The evaluation of RAG systems has become increasingly sophisticated. ARES \cite{saadf2024ares} provides an automated evaluation framework using synthetic training data and lightweight LM judges to assess context relevance, answer faithfulness, and answer relevance. A comprehensive study \cite{li2024raglongcontext} comparing RAG with long-context LLMs reveals that while long-context models achieve better average performance when sufficiently resourced, RAG maintains a significant cost advantage.

Privacy and security concerns have also emerged. Research on privacy in RAG \cite{zeng2024privacy} shows that RAG systems can leak their private retrieval databases. VersionRAG does not address this, although its graph records the source document and version of every chunk.

\begin{table*}[t]
\centering
\small
\caption{Comparison of RAG approaches for version-specific retrieval requirements. VersionRAG is the only method that directly addresses versioned documents, distinct from more generic temporal retrieval methods.}
\begin{tabular}{lcccccc}
\hline
\textbf{Method} & \makecell{Version\\Awareness} & \makecell{Change\\Tracking} & \makecell{Temporal\\Filtering} & \makecell{Graph\\Reasoning} & \makecell{Hybrid\\Retrieval} & \makecell{Efficient\\Indexing} \\
\hline
Standard RAG     & \xmark & \xmark & \xmark & \xmark & \xmark & \cmark \\
GraphRAG         & \xmark & \xmark & \xmark & \cmark & \cmark & \xmark \\
Time-Aware LM    & \xmark & \xmark & \cmark & \xmark & \xmark & N/A \\
CRONKGQA         & \xmark & \xmark & \cmark & \cmark & \xmark & \xmark \\
TempRALM         & \xmark & \xmark & \cmark & \xmark & \xmark & \cmark \\
TG-RAG           & \xmark & \xmark & \cmark & \cmark & \cmark & \xmark \\
\hline
\textbf{VersionRAG} & \cmark & \cmark & \xmark & \cmark & \cmark & \cmark \\
\hline
\end{tabular}
\label{tab:rag-comparison}
\end{table*}

\section{Implementation Details}
\label{app:implementation_details}

\subsection{Document Processing}

\paragraph{Input Formats.}
The system accepts PDF and Markdown input. PDFs are converted to Markdown to preserve structural information while enabling text processing.

\paragraph{Chunking.}
Documents are segmented line by line into chunks of at most 1,024 characters (about 250 tokens), split preferably at heading boundaries and never inside a code block; chunks do not overlap and no content is dropped.

\subsection{Indexing}

\paragraph{Attribute Extraction.}
Document metadata (topic title, short content summary, version) is obtained from the opening pages of each file. Where documents carry machine-readable version markers (version-bearing filenames or explicit \texttt{Version:} headers), extraction is a parse and requires no LLM call, with per-project normalization of the raw version suffix; where they do not, an LLM extractor supplies the same fields. At least the first ten pages are inspected to classify a document as changelog or general documentation, which determines whether change extraction reads explicit changelog entries or computes diffs.

\paragraph{Document Clustering.}
Documents are grouped into version lineages by title similarity and into categories by topic similarity.

\paragraph{Change Detection.}
Explicit changes are read from documents identified as changelogs. Implicit changes are computed with the DeepDiff library \cite{Dehpour_DeepDiff}, which performs line-level comparison between consecutive versions, followed by LLM interpretation of each diff into a natural-language change description attached to the corresponding Change node.

\subsection{Retrieval}

\paragraph{Query Classification.}
Queries are classified into three retrieval modes (\textsc{ContentRetrieval}, \textsc{VersionRetrieval} and \textsc{ChangeRetrieval}), together with any category, document and version parameters they carry.

\paragraph{Routing.}
\textsc{VersionRetrieval} resolves by graph traversal over Version nodes; \textsc{ChangeRetrieval} by lookup or semantic search over Change nodes; \textsc{ContentRetrieval} by vector search, with a version payload filter applied whenever the query carries a version constraint.

\paragraph{Prompting.}
All LLM subtasks (query classification with parameter extraction, indexing-time attribute
extraction, change interpretation, and answer generation) use instruction prompts with
in-prompt exemplars and no fine-tuning. Every call runs at temperature $0$. The three subtasks
that must return machine-readable output (classification/parameter extraction, attribute
extraction, change interpretation) are decoded under a JSON-schema constraint derived from a
Pydantic model, so out-of-vocabulary labels are impossible rather than merely discouraged;
answer generation is an unconstrained chat completion. Complete prompt texts are in
Appendix~\ref{app:prompts}, three worked end-to-end traces in Appendix~\ref{app:worked}, and
the models, endpoints, and library versions in Appendix~\ref{app:versions}.

\subsection{Query-Parser Routing Accuracy}
\label{app:parser}

Section~\ref{sub:routing} reports per-category routing accuracy
(Table~\ref{tab:parseraccuracy}) and the failure mode it isolates. This appendix gives the
per-backbone confusion matrices behind it. We score the parser's \texttt{query\_type} output
against the gold category of all 869 VersionQA-Software questions, once per backbone (within a
run, the same model classifies and generates). The two content categories, Content Retrieval and
Content Retrieval Complex, map onto the single classifier label \texttt{content\_retrieval}, so
the six gold categories of Table~\ref{tab:parseraccuracy} collapse to five rows in the confusion
matrices of Table~\ref{tab:parserconfusion}, with one content row of $150 + 139 = 289$ questions.

\begin{table*}[t]
\centering\small
\caption{Confusion matrices of the query parser on the 869 software questions, one block per
backbone. Rows are gold query types, columns parser predictions; diagonal entries in bold. The
two content categories share one classifier label and are therefore pooled into a single
\texttt{content} row.}
\begin{tabular}{llrrrrr}
\toprule
Backbone & gold $\backslash$ parsed & \texttt{chg (e)} & \texttt{chg (i)} & \texttt{content} & \texttt{ver-spec} & \texttt{ver-list} \\
\midrule
\multirow{5}{*}{\texttt{gpt-5.4-nano}}
 & \texttt{chg (e)}  & \textbf{116} & 5 & 2 & 26 & 0 \\
 & \texttt{chg (i)}  & 0 & \textbf{143} & 0 & 0 & 0 \\
 & \texttt{content}  & 0 & 0 & \textbf{282} & 7 & 0 \\
 & \texttt{ver-spec} & 1 & 1 & 1 & \textbf{140} & 0 \\
 & \texttt{ver-list} & 0 & 0 & 0 & 1 & \textbf{144} \\
\midrule
\multirow{5}{*}{\texttt{gpt-5.4-mini}}
 & \texttt{chg (e)}  & \textbf{105} & 29 & 3 & 12 & 0 \\
 & \texttt{chg (i)}  & 0 & \textbf{143} & 0 & 0 & 0 \\
 & \texttt{content}  & 0 & 0 & \textbf{289} & 0 & 0 \\
 & \texttt{ver-spec} & 0 & 0 & 5 & \textbf{138} & 0 \\
 & \texttt{ver-list} & 0 & 0 & 0 & 0 & \textbf{145} \\
\midrule
\multirow{5}{*}{Mistral Small 4}
 & \texttt{chg (e)}  & \textbf{106} & 17 & 2 & 24 & 0 \\
 & \texttt{chg (i)}  & 0 & \textbf{143} & 0 & 0 & 0 \\
 & \texttt{content}  & 0 & 0 & \textbf{287} & 2 & 0 \\
 & \texttt{ver-spec} & 0 & 1 & 0 & \textbf{142} & 0 \\
 & \texttt{ver-list} & 0 & 0 & 0 & 0 & \textbf{145} \\
\midrule
\multirow{5}{*}{\texttt{qwen3-32b}}
 & \texttt{chg (e)}  & \textbf{109} & 4 & 1 & 34 & 1 \\
 & \texttt{chg (i)}  & 0 & \textbf{143} & 0 & 0 & 0 \\
 & \texttt{content}  & 0 & 0 & \textbf{288} & 1 & 0 \\
 & \texttt{ver-spec} & 0 & 0 & 0 & \textbf{143} & 0 \\
 & \texttt{ver-list} & 0 & 0 & 0 & 0 & \textbf{145} \\
\bottomrule
\end{tabular}
\label{tab:parserconfusion}
\end{table*}

The matrices locate the error precisely. Explicit-change questions are grounded in
a changelog page and frequently name the version whose entry they ask about, which also satisfies
the version-specific rule; three of the four backbones therefore lose most of their explicit-change
mass to \texttt{ver-spec} (26, 24 and 34 questions). \texttt{gpt-5.4-mini} fails differently and
more often: it over-applies the implicit-change rule, sending 29 explicit-change questions to
\texttt{chg (i)}, which is why it is the weakest router (94.4\%) despite being the strongest
generator. Implicit change is never missed by any backbone: the rule is tested first and its
phrasing (``in which version did $X$ change'') is distinctive.

\subsection{Attribute Extraction: Parser vs.\ LLM}
\label{app:attrpath}

Attribute extraction is hybrid by design (Section~\ref{sec:versionrag}): documents carrying
machine-readable version markers are parsed at zero LLM cost, and an LLM extractor supplies the same fields where such markers are missing. Because both released corpora fall in the first case, the two paths can be
compared directly, holding everything else in the pipeline fixed. We index the software corpus twice,
once per path, and evaluate the full system on all 869 questions in both arms
(Table~\ref{tab:attrpath}).

\begin{table}[ht]
\centering\small
\caption{Attribute-extraction path, full VersionRAG on the software corpus ($n=869$). The two paths
are statistically equivalent within a $\pm 3$\,pp margin.}
\begin{tabular}{lcc}
\toprule
Extraction path & Acc. [\%] & 95\% Wilson CI \\
\midrule
LLM extractor          & 86.88 & [84.5--89.0] \\
Deterministic parser   & 87.57 & [85.2--89.6] \\
\bottomrule
\end{tabular}
\label{tab:attrpath}
\end{table}

The paired difference is $+0.69$\,pp in the parser's favor, over 811 concordant pairs and 58
discordant ones (32 parser-only-correct, 26 LLM-only-correct). An exact McNemar test does not reject
equality ($p=0.512$), and a 10k-resample bootstrap puts the 95\% interval on the paired difference at
$[-1.04, +2.42]$\,pp. A non-significant test only fails to find a difference, so we additionally run
two one-sided tests (TOST) for equivalence, which reject the presence of a difference larger than the
margin: $p=0.004$ at $\pm 3$\,pp and $p<0.0001$ at $\pm 5$\,pp. We therefore claim the paths
\emph{equivalent within $\pm 3$\,pp}, rather than claiming no difference exists.

\begin{table*}[t]
\centering\small
\caption{Per-category paired comparison of the two attribute-extraction paths, Holm-corrected across
the six categories. $b$ counts questions the parser arm answers correctly and the LLM arm does not,
$c$ the reverse; $\Delta = (b-c)/n$.}
\begin{tabular}{lrrrrrr}
\toprule
Category & $n$ & $b$ & $c$ & $\Delta$ [pp] & raw $p$ & Holm $p$ \\
\midrule
Content Retrieval Complex          & 139 & 8 & 0  & $+5.8$ & 0.0078 & 0.047\phantom{0} \\
Change Retrieval (implicit)        & 143 & 6 & 15 & $-6.3$ & 0.078\phantom{0} & 0.392 \\
Version Listing \& Inquiry         & 145 & 4 & 0  & $+2.8$ & 0.125\phantom{0} & 0.500 \\
Change Retrieval (explicit)        & 149 & 7 & 2  & $+3.4$ & 0.180\phantom{0} & 0.539 \\
Version-Specific Content Retrieval & 143 & 5 & 7  & $-1.4$ & 0.774\phantom{0} & 1.000 \\
Content Retrieval                  & 150 & 2 & 2  & $\pm0.0$ & 1.000\phantom{0} & 1.000 \\
\midrule
Overall                            & 869 & 32 & 26 & $+0.7$ & 0.512\phantom{0} & --- \\
\bottomrule
\end{tabular}
\label{tab:attrpathcat}
\end{table*}

We do not read Table~\ref{tab:attrpathcat} as parity in every category. Five of the six
categories are flat. One cell survives Holm correction, Content Retrieval Complex at $+5.8$\,pp
($p=0.047$), and it favors the \emph{parser}; the largest raw effect in the LLM extractor's favor,
$-6.3$\,pp on implicit change, does not survive correction. The two point in opposite directions and
cancel at the aggregate level, which is why the overall comparison is even. The defensible claims are
therefore aggregate equivalence within $\pm 3$\,pp, and that there is no category in which LLM
extraction significantly outperforms parsing. Our results therefore do not depend on documents having parsable version markers: using the LLM extractor everywhere costs no measurable accuracy.

\subsection{Baseline and Ablation Configurations}

\paragraph{Naive RAG.}
Dense retrieval with the same embedding model as VersionRAG, retrieving the top-$k$ chunks by cosine similarity with no metadata filtering or version awareness. This is a deliberately minimal configuration representing common production practice.

\paragraph{GraphRAG.}
Neo4j's GraphRAG package (\texttt{neo4j-graphrag}), using \texttt{SimpleKGPipeline} for indexing and \texttt{HybridCypherRetriever} for retrieval, which combines vector search with graph traversal and expands retrieved nodes along extracted relationships. Entities and relations are extracted by an LLM from every chunk at index time. The pipeline follows the Neo4j documentation and is not tuned for our corpus.

\paragraph{TG-RAG.}
The published implementation, which builds a bi-level temporal hierarchy (year\,$\to$\,quarter\,$\to$\,month) over timestamped relation quadruples extracted per chunk. We run its standard pipeline, untuned for our corpus, on the same corpus and with the same generator backbone as every other system in its column. We deliberately do not remap semantic versions onto synthetic timestamps to satisfy its normalizer. Such a remapping would supply from the outside the very version ordering the experiment is testing for, and it cannot be done faithfully in any case, because release dates do not order versions: Python 3.8.10 and 3.9.5 shipped on the same day, and documentation for an older minor line is frequently published later than for a newer one. We therefore report TG-RAG as evidence about the temporal axis, and not as an upper bound on what a version-adapted temporal graph could reach. That adaptation is not a parameter change: giving a temporal graph a canonical version order, version-scoped content boundaries and materialized inter-version change facts is the design we propose and evaluate here, so any gain it produced would be attributable to the version ontology rather than to the host method.

\paragraph{VersionRAG w/o Change.}
Identical to VersionRAG in attribute extraction, chunking, embeddings, graph construction, query routing, retrieval and generation, differing only in that no Change nodes are constructed.

\subsection{Resource Tracking}
\label{app:resources}

For every run we record input and output token counts, API cost, wall-clock time per phase, and the
size of the resulting graph in nodes and relationships. Table~\ref{tab:indexcost} reports the
index-time LLM cost of the three graph-building systems on the software corpus.

\begin{table*}[t]
\centering
\small
\caption{Index-time LLM tokens on the software corpus (882 document versions), by backbone.
Embedding tokens are excluded, since all three systems embed the corpus once at a cost far below LLM extraction. VersionRAG is broken into its three index-time calls. Baselines run with their standard pipelines, untuned for our corpus. The last two rows carry the point: what one additional indexed version costs
under each ontology, and the total index-time budget as a multiple of the 5.29\,M-token corpus text.}
\begin{tabular}{lrrrrrr}
\toprule
& \multicolumn{4}{c}{VersionRAG} & \multirow{2}{*}{GraphRAG} & \multirow{2}{*}{TG-RAG} \\
\cmidrule(r){2-5}
Backbone & attr. & chglog & diff & total & & \\
\midrule
\texttt{gpt-5.4-nano}  & 1.96\,M & 1.06\,M & 1.60\,M & 4.62\,M & 154\,M & 50.8\,M \\
\texttt{gpt-5.4-mini}  & 1.97\,M & 1.05\,M & 1.90\,M & 4.92\,M & 154\,M & 43.7\,M \\
Mistral Small 4        & 2.13\,M & 1.18\,M & 1.65\,M & 4.96\,M & 119\,M & 43.0\,M \\
\texttt{qwen3-32b}     & 2.17\,M & 1.13\,M & 1.74\,M & 5.04\,M & 143\,M & 55.5\,M \\
\midrule
Per indexed version & \multicolumn{4}{c}{5.2--5.7\,K} & 135--175\,K & 49--63\,K \\
$\times$ corpus size & \multicolumn{4}{c}{0.87--0.95} & 22--29 & 8--10 \\
\bottomrule
\end{tabular}
\label{tab:indexcost}
\end{table*}

VersionRAG issues 2,195--2,199 index-time calls, namely one attribute-extraction call per document
version plus one change-extraction call per consecutive version pair (882 + 1{,}316), routed to the
changelog or the file-diff path according to the document's bucket. That is 5{,}200--5{,}700 tokens
per indexed document version, of which 0.70--1.03\,M tokens per run are generated output. The
difference in magnitude follows from what is extracted, not from prompt efficiency, and it is
specific to versioned corpora. Such a corpus is largely self-overlapping, since consecutive versions
of a document share most of their text, so per-chunk extraction re-derives nearly the same entities
or quadruples once per version: the cost grows with the number of versions even though the
information each version contributes does not. VersionRAG instead pays once per version for
attributes and once per consecutive pair for the change between them, which is the quantity that
actually grows with the corpus.

The clearest way to see the difference is to normalize by the corpus itself, which is 5.29\,M tokens
of text (last row of Table~\ref{tab:indexcost}). VersionRAG's complete index-time LLM budget is
$0.87$--$0.95\times$ the corpus, that is, less than one pass over it: attribute extraction reads only
the head of each file and change extraction reads only changelog entries or diffs, so no stage ever
reads every version in full. GraphRAG spends $22$--$29\times$ the corpus and TG-RAG $8$--$10\times$,
since per-chunk extraction sends every chunk of every version through an instruction prompt and
generates output for each. Embeddings, excluded throughout, do read the corpus once for all three
systems, at a unit cost roughly two orders of magnitude below generation.

Three qualifications. First, this is a statement about the ontology rather than about the two
implementations, both of which we run with their standard pipelines, untuned for our corpus; efficiency-oriented variants of
the same design, such as LightRAG \cite{guo2024lightrag} and HippoRAG
\cite{gutierrez2024hipporag}, lower the constant but not the scaling, because they still extract per
chunk over a flat entity--relation graph. Second, VersionRAG's index is deliberately narrower: it
builds no entity--relation graph, so it cannot serve the entity-centric multi-hop queries GraphRAG
supports, and the saving reflects that scope difference as well as the change-oriented design. The
comparison is therefore about indexing versioned corpora for the query types of
Section~\ref{sub:problem} and is not a general efficiency claim about either system. Third, we
report tokens rather than currency, because per-token prices differ across the four backbones and
across routes for the same model, so a cost ratio would depend on procurement rather than on the
systems.

\subsection{Prompts}
\label{app:prompts}

VersionRAG issues five prompts (Table~\ref{tab:prompts}), all instruction prompts with in-prompt
exemplars and no fine-tuning. The full texts are released with the code (\githuburl); four design
decisions carry the paper's claims.

\begin{table*}[t]
\centering\footnotesize
\caption{The five prompts, their stage, and their output contract. Schema-constrained calls are
decoded against a Pydantic model, so malformed or out-of-vocabulary output is impossible rather
than merely discouraged.}
\label{tab:prompts}
\begin{tabular}{@{}llp{0.21\textwidth}p{0.31\textwidth}@{}}
\toprule
\textbf{Prompt} & \textbf{Stage} & \textbf{Input} & \textbf{Output} \\
\midrule
Attribute extraction & indexing & file path + first 2\,000 chars & \texttt{\{title, version, doc\_type, category\_hint, confidence\}} \\
\addlinespace
Change interpretation & indexing & changelog text or diff JSON & list of \texttt{\{name, description\}} \\
\addlinespace
Classification + parameters & query & question + document catalog & \texttt{\{query\_type, document, version, version\_from, version\_to, category\}} \\
\addlinespace
Answer generation & query & retrieved context + question & free-form answer \\
\addlinespace
Judge & evaluation & question, reference, answer & \texttt{1} or \texttt{0} \\
\bottomrule
\end{tabular}
\end{table*}

First, the three schema-constrained subtasks are decoded against a Pydantic model, so
\texttt{query\_type} and \texttt{category\_hint} cannot take values outside the taxonomy or the corpus category set. This prevents one document from being filed under two categories, which would split its version lineage, already during decoding instead of detecting it afterwards. Second, the classifier's five
rules are applied in a fixed order with first match winning, and the implicit-change rule is
tested \emph{before} the version-specific rule: naming a function or option does not make a
question version-specific, so ``in which version was \texttt{sync.Map.Clear} added'' routes to
change retrieval, where the version is the answer, rather than to content retrieval, where it
would become a filter. The extraction rules further forbid populating \texttt{version\_from} or
\texttt{version\_to} for implicit-change queries. Third, the classifier prompt carries a
rendering of the documents currently in the graph and requires \texttt{document} and
\texttt{category} to match it exactly, so document identity is resolved against the index rather
than generated, which is what lets retrieval use exact-match payload filters instead of text
matching. Fourth, change interpretation is selected by \texttt{doc\_type}: changelog documents
route to a variant that reads release-note text directly, documentation documents to one that
reads a \texttt{DeepDiff} between consecutive versions, so the explicit/implicit split follows
the source partition rather than a judgment at generation time.

The legal corpus swaps three prompts. Versions are ISO dates, so extraction normalizes every
human date form to \texttt{YYYY-MM-DD}, and \texttt{change\_retrieval\_explicit} is redefined as
``what changed between two dated versions'' and requires both endpoints, because UK Acts have no
changelog documents. Generation and judge prompts are identical across corpora.

\subsection{Worked End-to-End Examples}
\label{app:worked}

The classifier emits five labels which the router collapses onto three retrieval paths: version
listing and both change classes go to the graph, both content classes to the vector store. We
trace one query per path. All three are real items from the software evaluation set, and the
reported label, retrieval mode, and latency are those logged during the main run
(\texttt{gpt-5.4-mini} backbone, \texttt{text-embedding-3-small} index).

\paragraph{Path 1: version listing (graph, Version nodes).}
\emph{``What is the latest numpy version you know of?''}
The first classification rule fires, since the question is about the set of versions, not
about content, giving \texttt{query\_type = version\_listing}, \texttt{document = "numpy"},
\texttt{category = "numpy"}, and all three version fields \texttt{null}. The document name is
not a guess: the injected catalog contains exactly one \texttt{"numpy"} entry, because the
attribute extractor mapped all 21 files under \texttt{changelog/numpy\_releases/} onto that
title. Routing goes to \texttt{GraphRetriever.retrieve\_versions}; the vector store is not
touched. With a document in hand the traversal is one hop, ordered by the integer sort key
computed at indexing time:

\begin{codeblock}\begin{verbatim}
MATCH (doc:Document {name: $document})
      -[:HAS_VERSION]->(v:Version)
RETURN v.number AS number,
       v.sort_key AS sort_key,
       v.doc_type AS doc_type
ORDER BY v.sort_key
\end{verbatim}\end{codeblock}

\begin{figure*}[t]
    \centering
    \includegraphics[width=\textwidth]{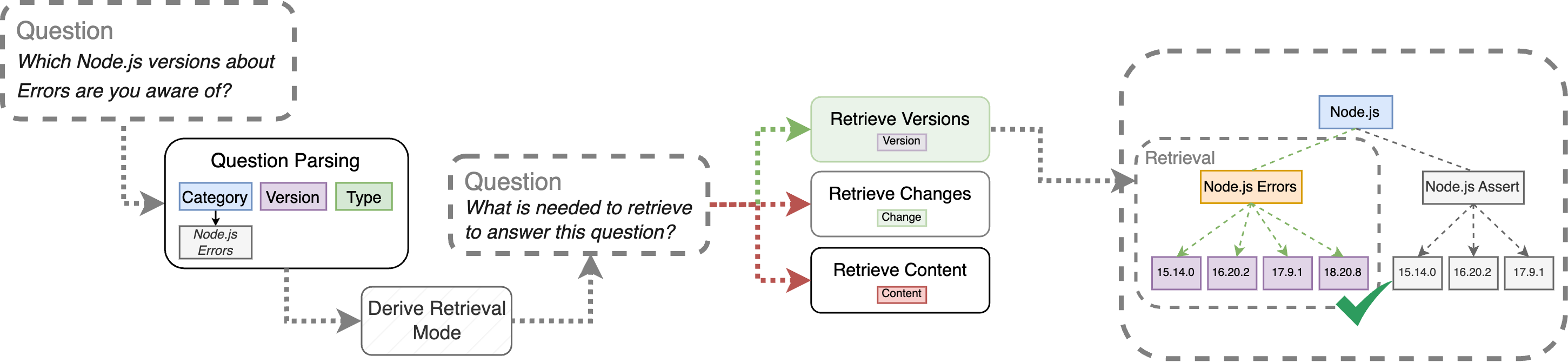}
    \caption{Path 1, version listing. The query resolves by traversal over the Version nodes of
    the addressed document; the sibling document's versions are reachable but out of scope.}
    \label{fig:trace-version}
\end{figure*}

\noindent All 21 versions return, \texttt{2.0.0} through \texttt{2.4.0}. Sort keys are numeric
(\texttt{2.4.0} $\rightarrow$ 20\,400, \texttt{2.10.0} $\rightarrow$ 21\,000), so ``latest'' is
a total order over release numbers rather than lexicographic order over strings, which would
rank \texttt{2.10.0} below \texttt{2.4.0}. The rows are serialized one per line and paired with
the version-listing generation prompt, yielding \emph{``The latest numpy version listed is 2.4.0''}
against the reference \emph{``The latest numpy version is 2.4.0''}; judge 1, latency
1.58\,s. Figure~\ref{fig:trace-version} shows the same path on a documentation lineage. No
system that retrieves by content similarity can answer this reliably, because the answer is a
property of the corpus rather than a statement inside any chunk.

\begin{figure*}[t]
    \centering
    \includegraphics[width=\textwidth]{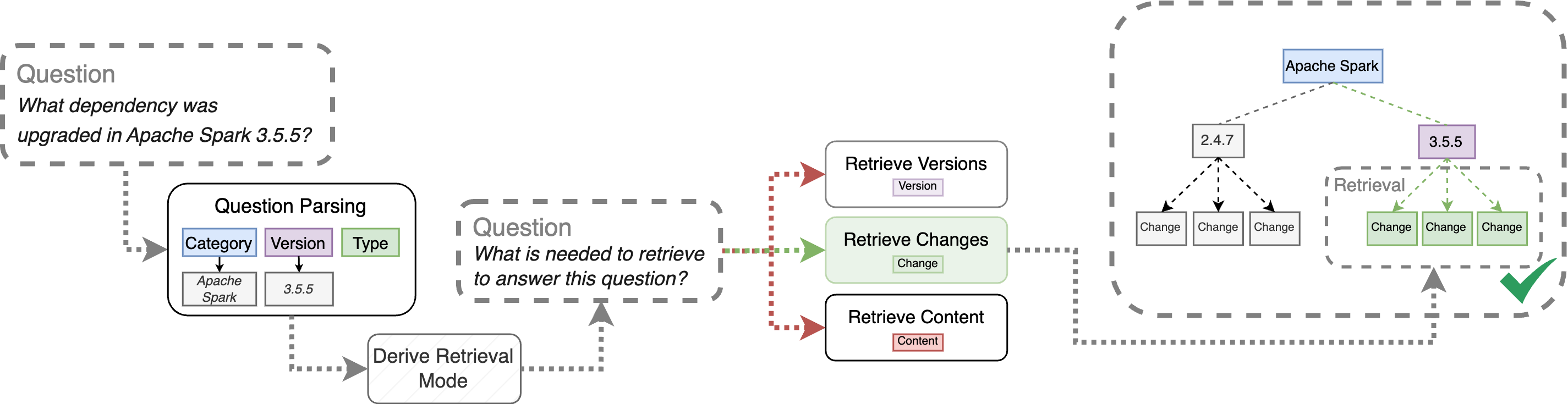}
    \caption{Path 2, change retrieval. The router bypasses vector search and reads the Change
    nodes attached to the addressed version. The document shown is illustrative and is not part
    of the released corpus.}
    \label{fig:trace-change}
\end{figure*}

\paragraph{Path 2: implicit change retrieval (graph, Change nodes).}
\emph{``In which Go version was the \texttt{Map.Clear} method added to the \texttt{sync}
package?''}
The implicit-change rule fires. The prompt states explicitly that naming an identifier such as \texttt{sync.Map.Clear} does \emph{not} pin a version, because here the version is the answer, not a constraint. The parser therefore returns \texttt{change\_retrieval\_implicit}, \texttt{document =
"golang\_sync"}, \texttt{category = "golang"}, and all version fields \texttt{null}. Pinning a version would restrict retrieval to the version the model already suspects. Routing goes
to \texttt{GraphRetriever.retrieve\_changes} (Figure~\ref{fig:trace-change}). Since the document is known and no version is given, the retriever takes the full-text branch over Change nodes, scoped to
the document:

\begin{codeblock}\begin{verbatim}
CALL db.index.fulltext.queryNodes(
       "change_fulltext_idx", $search_term)
YIELD node, score
MATCH (v1)-[:CHANGED_TO]->(node)
      <-[:CHANGED_FROM]-(v2)
MATCH (cat:Category)-[:HAS_DOC]->
      (d:Document {name: v1.document})
WHERE ($document IS NULL
       OR v1.document = $document)
  AND ($category IS NULL
       OR cat.name = $category)
RETURN node.name, node.description,
       node.change_type, node.source,
       v1.number AS version_from,
       v2.number AS version_to, score
ORDER BY score DESC LIMIT 40
\end{verbatim}\end{codeblock}

\noindent The document filter is what makes this tractable: unscoped, \texttt{"Clear"} and
\texttt{"Map"} match thousands of Change nodes. The $\leq$40 candidates are re-ranked by cosine
similarity between the question embedding and each candidate's \texttt{name~--~description}
string, keeping the top 8; Lucene tf--idf alone ranks by term overlap and is too unscoped for
identifier queries, and the semantic re-rank is what promotes \texttt{Map.Clear} over sibling
\texttt{Map} methods.

Because \texttt{golang\_sync} is a documentation document, its Change nodes were built by the
implicit path. For the \texttt{1.22.0} $\rightarrow$ \texttt{1.23.0} pair, \texttt{DeepDiff}
over the line-split files yields 193 entries (23.9\,kB of JSON); noise filtering drops one and
entry-aware chunking splits the rest into 5 chunks, each sent to the implicit change
prompt in parallel. Two of the four answer-bearing entries:

\begin{codeblock}\begin{verbatim}
{"iterable_item_added": {"root[342]":
   "#### func (\\*Map) [Clear](https://cs.
   opensource.google/go/go/+/go1.23.0:src/
   sync/map.go;l=162) added in go1.23.0"}}
{"iterable_item_added": {"root[348]":
   "Clear deletes all the entries,
   resulting in an empty Map."}}
\end{verbatim}\end{codeblock}

\noindent The extractor turns these into a single Change node naming \texttt{(*Map) Clear},
with \texttt{change\_type = implicit} and \texttt{source = file\_diff}, wired as
\texttt{(v:1.22.0)-[:CHANGED\_TO]->}\allowbreak\texttt{(c)<-[:CHANGED\_FROM]-(v:1.23.0)}. The version endpoints are
structural, so the generator reads \texttt{version\_to} off the edge instead of parsing a
version out of prose: \emph{``The \texttt{sync.Map.Clear} method was added in Go 1.23.0''};
judge 1, latency 2.50\,s. This is the path that separates VersionRAG from its ablation, which
has the same graph, catalog, and version filters but no change layer, and scores 27 points
lower on implicit change (Table~\ref{tab:capability}).

\paragraph{Path 3: version-specific content retrieval (vector, filtered).}
\emph{``What is the default value of the \texttt{dim\_feedforward} parameter in the
\texttt{torch.nn.Transformer} class in PyTorch version 2.5?''}
The version-specific rule fires: a version is pinned and the question asks what exists in it. Note the contrast
with Path~2: same corpus, same kind of identifier, opposite role for the version. The parser
returns \texttt{content\_version\_specific}, \texttt{document = "pytorch\_transformer"},
\texttt{version = "2.5"}, \texttt{category = "pytorch"}; the version string is normalized so it
matches the payload value written at indexing time from the body line \texttt{Version: 2.5}.
Because the label is version-specific \emph{and} a version was extracted, a hard payload filter
is emitted, and the constraint is enforced by the store, not suggested to the ranker:

\begin{codeblock}\begin{verbatim}
Filter(must=[
  FieldCondition(key="version",
    match=MatchValue(value="2.5")),
  FieldCondition(key="document",
    match=MatchValue(
      value="pytorch_transformer")),
  FieldCondition(key="category",
    match=MatchValue(value="pytorch")),
])
\end{verbatim}\end{codeblock}

\noindent \texttt{document} uses exact \texttt{MatchValue} rather than text matching, so
\texttt{pytorch\_transformer} cannot bleed into sibling documents. Up to 15 chunks are
retrieved under that filter by dense similarity. The top chunk is the class signature from
\texttt{docs/pytorch/transformer/}\allowbreak\texttt{transformer\_25.md}:

\begin{codeblock}\begin{verbatim}
--- Document: pytorch_transformer,
    Version: 2.5 ---
*class* torch.nn.Transformer(
  *d\_model=512*, *nhead=8*,
  *num\_encoder\_layers=6*,
  *num\_decoder\_layers=6*,
  *dim\_feedforward=2048*, *dropout=0.1*,
  *activation=<function relu>*, ...)
\end{verbatim}\end{codeblock}

\begin{figure*}[t]
    \centering
    \includegraphics[width=\textwidth]{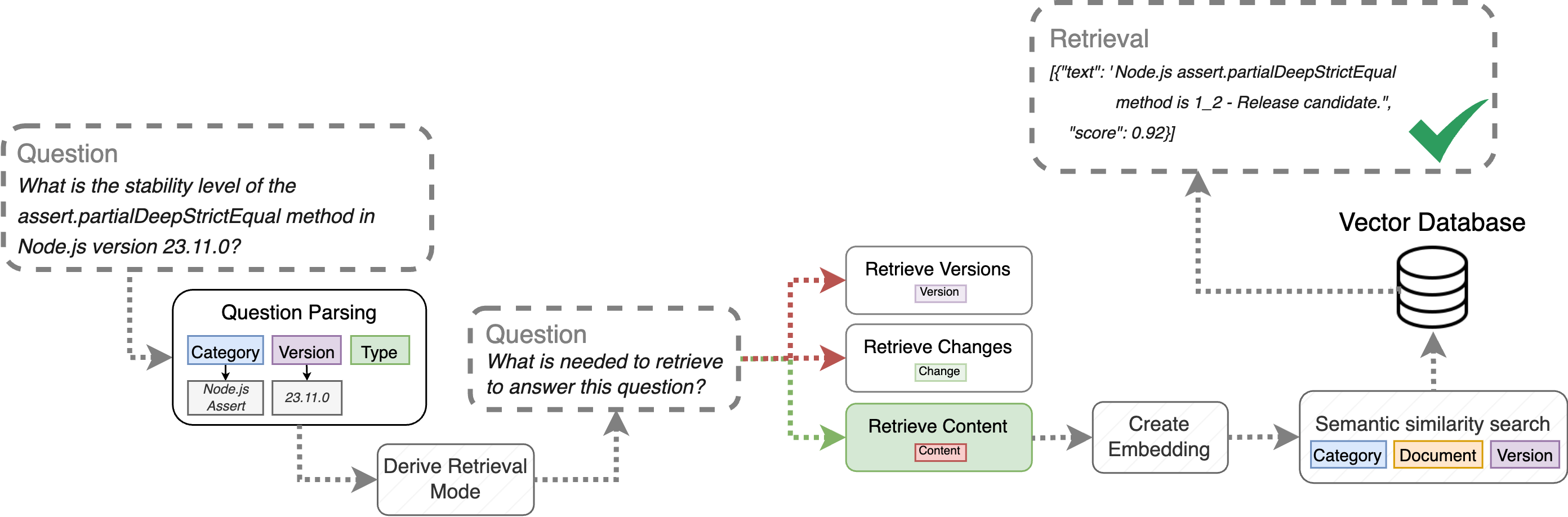}
    \caption{Path 3, version-specific content retrieval. The extracted version constraint
    becomes a hard payload filter, so only chunks of the named version are admissible.}
    \label{fig:trace-content}
\end{figure*}

\begin{figure*}[t]
    \centering
    \includegraphics[width=0.75\textwidth]{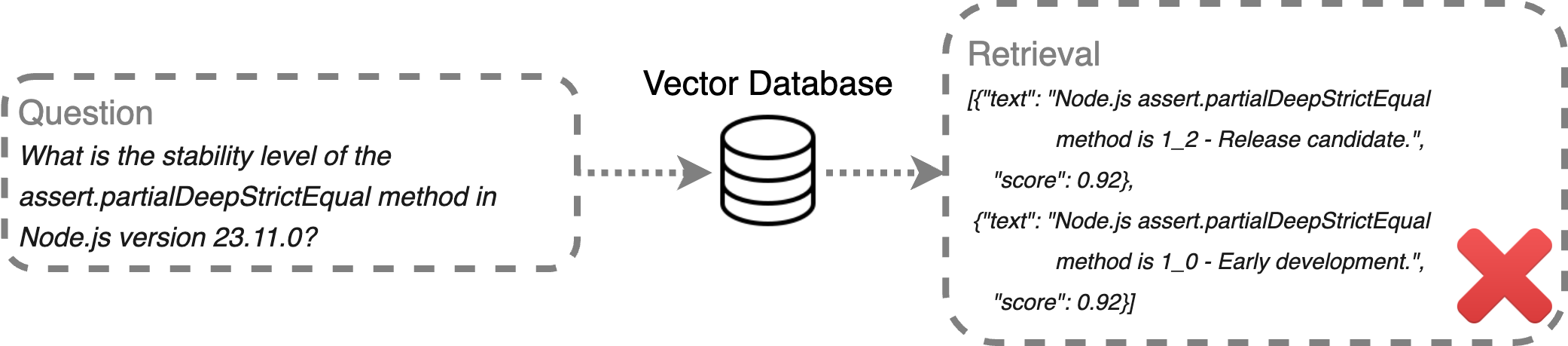}
    \caption{A version-specific query under flat retrieval. Two versions of the same method tie
    at 0.92 with contradictory answers, which is version conflation in its minimal form, since the
    similarity function is effectively version-invariant.}
    \label{fig:trace-naive}
\end{figure*}

\noindent Six other versions of this document (2.0--2.6) contain the same parameter list in the
same surrounding prose; the payload filter is the only thing keeping their chunks out of the
context window. Answer: \emph{``The default value of \texttt{dim\_feedforward} in
\texttt{torch.nn.Transformer} for PyTorch 2.5 is 2048''}; judge 1, latency 2.10\,s.
Figures~\ref{fig:trace-content} and~\ref{fig:trace-naive} contrast this path with flat
retrieval on a Node.js query, where two chunks from different versions tie at 0.92 with
contradictory answers.

\paragraph{The remaining two labels} reuse this machinery unmodified.
\texttt{content\_retrieval} runs the same vector path with the version condition omitted
(document and category filters still apply when extracted), which is why the two content
classes share a generation prompt. \texttt{change\_retrieval\_explicit} runs the same graph path
as Path~2, except that when the question names a version the retriever takes the exact-edge
branch instead of full-text search, so no ranking step is involved, since the edges already encode
both endpoints. If a change query returns no Change nodes on any branch, the router falls back
to vector retrieval, so the query degrades to naive-RAG behavior rather than returning nothing.

\subsection{Models, Endpoints, and Versions}
\label{app:versions}

Each backbone in the results tables is a single substitution: the same model serves
classification, generation, and both indexing-time extraction subtasks, so a row measures one
model across the whole pipeline (Table~\ref{tab:model-roles}). The judge is held fixed at
\texttt{gpt-5.4} for every row, and every run embeds with \texttt{text-embedding-3-small}.
Table~\ref{tab:endpoints} gives the serving endpoint and decoding settings behind each backbone.

\begin{table*}[t]
\centering\small
\caption{LLM and embedding assignment per subtask. \emph{Backbone} is the model named in the
results tables; the four backbone rows are set together.}
\label{tab:model-roles}
\setlength{\tabcolsep}{4pt}
\begin{tabular}{llll}
\toprule
\textbf{Subtask} & \textbf{Setting} & \textbf{Model} & \textbf{Decoding} \\
\midrule
Classification + parameters & \texttt{LLM\_MODEL\_FAST} & backbone & schema-constrained \\
Answer generation & \texttt{LLM\_MODEL\_STANDARD} & backbone & free-form chat \\
Attribute extraction (index) & \texttt{LLM\_MODEL\_EXTRACT} & backbone & schema-constrained \\
Change interpretation (index) & \texttt{LLM\_MODEL\_CHANGE} & backbone & schema-constrained \\
Scoring (judge) & \texttt{LLM\_MODEL\_STRONG} & \texttt{gpt-5.4} (fixed) & free-form (1 token) \\
\midrule
Embeddings & \texttt{EMBEDDING\_MODEL} & \texttt{text-embedding-3-small} (1536-d, dense) & --- \\
\bottomrule
\end{tabular}
\end{table*}

\begin{table*}[t]
\centering\small
\caption{Backbones and serving endpoints. All calls use temperature $0$ (omitted for
\texttt{gpt-5.4-nano}, which rejects the parameter) and cap output at 12\,000 tokens.}
\label{tab:endpoints}
\begin{tabular}{lll}
\toprule
\textbf{Backbone (as reported)} & \textbf{Provider} & \textbf{Endpoint} \\
\midrule
\texttt{gpt-5.4-nano} & OpenAI & \texttt{https://api.openai.com/v1} \\
\texttt{gpt-5.4-mini} & OpenAI & \texttt{https://api.openai.com/v1} \\
\texttt{mistralai/mistral-small-2603} & OpenRouter & \texttt{https://openrouter.ai/api/v1} \\
\texttt{qwen/qwen3-32b} & OpenRouter & \texttt{https://openrouter.ai/api/v1} \\
\midrule
judge \texttt{gpt-5.4} & OpenAI & \texttt{https://api.openai.com/v1} \\
\texttt{text-embedding-3-small} & OpenAI & \texttt{.../v1/embeddings} \\
\bottomrule
\end{tabular}
\end{table*}

Routing is by model identifier: a \texttt{/} in the name selects the OpenRouter client,
anything else the direct OpenAI client. The two paths differ in three API details that affect
reproducibility. Structured output is native JSON-schema decoding on OpenAI but tool-calling
mode on OpenRouter, because at least one provider rejects
\texttt{response\_format=json\_object} unless the word \emph{json} appears in the messages. The
output cap is \texttt{max\_completion\_tokens} on OpenAI's GPT-5.x models and the legacy
\texttt{max\_tokens} on OpenRouter providers; the client sends whichever the route accepts.
OpenRouter calls set \texttt{provider.require\_parameters=true} so requests are only routed to
providers supporting every parameter sent, and are capped at 10 in-flight requests.
Schema-constrained calls retry up to 3 times on a validation failure; embedding calls retry up
to 8 times with exponential backoff.
Table~\ref{tab:tool-versions} records the service and library versions used for every run.

\begin{table*}[t]
\centering\small
\caption{Service and library versions. Python 3.12.}
\label{tab:tool-versions}
\begin{tabular}{lll}
\toprule
\textbf{Component} & \textbf{Version} & \textbf{Role} \\
\midrule
Neo4j (\texttt{neo4j:5-community}) & 5.x & version graph, Lucene full-text index \\
\texttt{neo4j} Python driver & 6.1.0 & async Bolt client \\
Qdrant (\texttt{qdrant/qdrant}) & v1.17.0 & chunk vectors + payload filters \\
\texttt{qdrant-client} & 1.17.1 & REST/gRPC client \\
\texttt{openai} & 2.30.0 & OpenAI- and OpenRouter-compatible client \\
\texttt{instructor} & 1.15.1 & schema-constrained decoding, retries \\
\texttt{pydantic} & 2.12.5 & response schemas \\
\texttt{deepdiff} & 9.0.0 & line-level document diffs \\
\texttt{pymupdf4llm} & 1.27.2.2 & PDF to Markdown conversion \\
\texttt{fastapi} / \texttt{uvicorn} & 0.135.3 / 0.42.0 & serving layer \\
\bottomrule
\end{tabular}
\end{table*}

Neo4j is reached over Bolt at \texttt{bolt://localhost:7687} and Qdrant over HTTP at
\texttt{localhost:6333}; both run from the bundled \texttt{docker-compose.yml}. The system is
served by \texttt{uvicorn} and exposes \texttt{POST /query} (parse, retrieve, generate),
\texttt{POST /query/retrieve-only} for retrieval-side diagnostics,
\texttt{POST /indexing/run} and \texttt{/indexing/add}, \texttt{GET /indexing/status}, and
\texttt{GET /health}. Evaluation drives \texttt{POST /query} once per question.
Table~\ref{tab:hyperparams} lists the hyper-parameters in force for every reported run.

\begin{table*}[t]
\centering\small
\caption{Hyper-parameters in force for the reported runs.}
\label{tab:hyperparams}
\begin{tabular}{lll}
\toprule
\textbf{Stage} & \textbf{Parameter} & \textbf{Value} \\
\midrule
Chunking (indexing and release) & strategy & lossless line-based, fence-aware \\
 & min / soft-max / hard-max & 250 / 800 / 1024 characters \\
\midrule
Attribute extraction & document prefix sent to LLM & 2\,000 characters \\
\midrule
Change interpretation & changelog text budget & 6\,000 characters \\
 & diff chunk budget & 40 entries or 8\,000 characters \\
 & noise filter & whitespace-only lines; digit/punctuation-only edits \\
\midrule
Vector retrieval & chunks returned & 15 \\
 & search & dense, under payload filter \\
\midrule
Graph retrieval & exact-edge result cap & 50 \\
 & full-text candidate cap & 40 \\
 & kept after semantic re-rank & 8 \\
\midrule
All LLM calls & temperature & 0.0 \\
 & reasoning & none \\
 & max output tokens & 12\,000 \\
 & schema-validation retries & 3 \\
\bottomrule
\end{tabular}
\end{table*}

\section{Paired Change-Layer Effects}
\label{app:fixbreak}

Table~\ref{tab:fixbreak} gives the per-backbone counts behind the aggregate quoted in
\S\ref{sub:mechanism}: for implicit-change questions, how many the change layer turns from
wrong to right (fixes) against how many it turns from right to wrong (regressions), paired at
the question level between VersionRAG and its ablation. The layer fixes far more than it
breaks on every backbone, and the direction is stable across generator families.

\begin{table}[ht]
\centering
\small
\caption{Implicit-change questions the change layer fixes versus breaks, paired per question
(VersionRAG w/o Change $\to$ VersionRAG, VersionQA-Software).}
\begin{tabular}{lrrr}
\toprule
Generator & Fixes & Regressions & Ratio \\
\midrule
\texttt{gpt-5.4-nano} & 35 & 3 & 11.7$\times$ \\
\texttt{gpt-5.4-mini} & 36 & 4 & 9.0$\times$ \\
Mistral Small 4 & 52 & 9 & 5.8$\times$ \\
\texttt{qwen3-32b} & 28 & 7 & 4.0$\times$ \\
\midrule
Total & 151 & 23 & 6.6$\times$ \\
\bottomrule
\end{tabular}
\label{tab:fixbreak}
\end{table}

\section{Judge Audit Protocol}
\label{app:judgeaudit}

The 306 software answers were hand-adjudicated against independently
human-verified references, blind to both automatic verdicts and system identity, under a
fixed coding rule and with zero abstentions. The frame is the full software population of
17,377 scored answers, split into four cells by judge verdict crossed with NLI
corroboration and allocated proportionally across the five systems. Cells are not sampled in proportion to their size: we sampled more heavily where errors are more likely and weigh more on the error rate. Weighting each cell by its population share when combining keeps the overall estimate unbiased.
Cell sizes targeted $\pm2$ pp on the false-positive rate and $\pm5$ pp on the
false-negative rate, achieving $\pm2.3$ and $\pm4.8$. Four further verdicts were dropped
after adjudication because the responses they scored were superseded by a re-run of the
GraphRAG$\times$\texttt{qwen3-32b} cell.

\begin{table*}[t]
\centering
\small
\caption{Judge error rates over the software population. Rate = fraction of adjudicated
verdicts in which the human adjudicator disagreed with the judge. False positive is
P(answer wrong $\mid$ judge accepted) and false negative P(answer right $\mid$ judge
rejected); each is the population-weighted combination of its two cells.}
\begin{tabular}{llrrrrr@{\quad}l}
\toprule
Cell & Definition & Population & Weight & $n$ & Errors & Rate & 95\% CI \\
\midrule
\multicolumn{8}{l}{\textbf{False positive} $=$ \textbf{2.5\% [0.2, 4.8]}} \\
A & judge $=1$, NLI $=0$ & 4,521 & 0.461 & 166 & 9 & 5.4\% & [2.9, 10.0] \\
C & judge $=1$, NLI $=1$ & 5,283 & 0.539 & 63 & 0 & 0.0\% & [0.0, 5.7] \\
\midrule
\multicolumn{8}{l}{\textbf{False negative} $=$ \textbf{5.5\% [0.8, 10.3]}} \\
B & judge $=0$, NLI $=1$ & 405 & 0.053 & 38 & 22 & 57.9\% & [42.2, 72.1] \\
D & judge $=0$, NLI $=0$ & 7,168 & 0.947 & 39 & 1 & 2.6\% & [0.5, 13.2] \\
\bottomrule
\end{tabular}
\label{tab:auditcells}
\end{table*}

Cell B carries the highest error rate in the audit: when the judge rejects an answer that
NLI accepts, it is wrong 57.9\% [42.2, 72.1] of the time. That stratum is only 5.3\% of
rejected answers, yet it supplies 3.1 of the 5.5 pp false-negative rate, the remaining
2.4 pp coming from the far larger cell D. False negatives are therefore concentrated in a
stratum NLI identifies in advance; false positives are not, and cell C, where both
metrics agree the answer is correct, contains no errors in 63 verdicts.

We release the audit sample and per-item verdicts, the cell and system key, the coding
rule, the sampling and analysis script (\texttt{--analyze} regenerates
Table~\ref{tab:auditcells}), and the NLI scores from which the strata were drawn
(\githuburl).

\section{Dataset Construction and Examples}
\label{app:datasets}

\subsection{Software Corpus}

The 882 software files are partitioned into a documentation bucket and a changelog bucket
(Table~\ref{tab:swprojects}). The partition is what defines the explicit/implicit change
split: explicit-change questions are drawn only from changelog files, implicit-change
questions only from diffs between consecutive documentation versions.

\begin{table}[ht]
\centering
\small
\caption{Software projects by bucket. Documentation: 313 files; changelogs: 569 files.
pandas appears in both.}
\begin{tabular}{ll}
\toprule
Bucket & Projects \\
\midrule
Documentation & Python, Node.js, Go, PyTorch, \\
(313 files) & pandas, Kubernetes, ROS~2, \\
 & NVIDIA TensorRT-LLM \\
\midrule
Changelogs & Docker Compose, Docker Engine, \\
(569 files) & Java JDK, MySQL, NumPy, pandas, \\
 & PostgreSQL, SciPy, TensorFlow, \\
 & Wireshark \\
\bottomrule
\end{tabular}
\label{tab:swprojects}
\end{table}

All software documentation and changelogs are publicly published project
artifacts, collected from official project documentation sites and release-note
pages and used here for research purposes consistent with their terms. The
projects in Table~\ref{tab:swprojects} publish under a mix of permissive
open-source and open-content licenses (BSD-3-Clause, MIT, Apache-2.0, the PSF
License, CC~BY~4.0, the PostgreSQL Licence, and GPLv2 for Wireshark);
per-file source URLs and the governing license are released
with the corpus in \texttt{LICENSES.md}.
Each project's documentation and release notes were retrieved from its
official documentation site or repository; per-project source sites are listed in \texttt{SOURCES.md} in the released corpus.

\textbf{Personal Data.} The corpus contains no user-generated or private content; the only personal
references are contributor credits in release notes, already published by the
projects in a professional capacity, which we retain unaltered as attribution.

\subsection{Legal Corpus}

Table~\ref{tab:legalacts} gives the composition of the legal corpus. Versions are the
distinct \texttt{RestrictStartDate} values within a Part, i.e., the in-force axis rather
than a publication date. Parts with fewer than three version dates are discarded (so every
retained Part has at least two transitions), Parts outside 5--250\,KB after Markdown
conversion are dropped, and each surviving Part is capped at four versions (oldest, newest,
and evenly spaced middles), which prevents the Income Tax Act from dominating the corpus.
Both ordinary sections and schedule Parts are retained as retrieval units.

\begin{table}[ht]
\centering
\small
\caption{VersionQA-Legal composition: five UK Public General Acts, 355 document versions
over 96 Parts.}
\begin{tabular}{lrr}
\toprule
Act & Files & Parts \\
\midrule
\makecell[l]{Income Tax (Earnings and\\Pensions) Act 2003} & 137 & 37 \\
Equality Act 2010 & 74 & 21 \\
Finance Act 2003 & 67 & 18 \\
Data Protection Act 2018 & 45 & 12 \\
Localism Act 2011 & 32 & 8 \\
\midrule
Total & 355 & 96 \\
\bottomrule
\end{tabular}
\label{tab:legalacts}
\end{table}

All legal source texts are Crown copyright material obtained from
legislation.gov.uk and are made available under the Open Government Licence
v3.0, which permits redistribution and adaptation with attribution; our derived
Markdown conversions and question--answer pairs are released under the same
attribution terms. Legislation text carries no personal data.

\subsection{Question Categories}

\begin{table}[ht]
\centering
\small
\caption{Question distribution per category. VersionQA-Software (SW) and VersionQA-Legal (LG) share the same six-category taxonomy; two thirds of the categories require version-aware reasoning.}
\setlength{\tabcolsep}{4pt}
\begin{tabular}{lccc}
\hline
Category & SW & LG & \makecell{Version-\\Sensitive} \\ \hline
Content Retrieval & 150 & 93 & No \\
Content Retrieval Complex & 139 & 86 & No \\
Content Retrieval Version-Specific & 143 & 97 & Yes \\
Version Listing \& Inquiry & 145 & 95 & Yes \\
Change Retrieval (Explicit) & 149 & 91 & Yes \\
Change Retrieval (Implicit) & 143 & 85 & Yes \\ \hline
Total & 869 & 547 & \\
\hline
\end{tabular}
\label{tab:querycategories}
\end{table}

Table~\ref{tab:catdefs} gives the operational definition used by the generator for each of
the six categories. Both datasets use the same taxonomy. The two categories marked not version-sensitive in Table~\ref{tab:querycategories} are version-agnostic only in their \emph{wording}: the question names no version, but it is generated from a single file, i.e., from one version's text. We do not check that the answer holds in every version of the document.

\begin{table*}[t]
\centering
\small
\caption{Operational category definitions used at generation time.}
\begin{tabular}{p{0.22\textwidth}p{0.72\textwidth}}
\toprule
Category & Definition \\
\midrule
Content Retrieval & Version-agnostic factual question grounded in a single documentation file. \\
Content Retrieval Complex & Multi-step question combining or reasoning over several passages of the \emph{same} file. \\
Content Retrieval Version-Specific & Names a version; the answer reflects that version. With $p=0.3$ the generator instead produces a feature-absence item over two versions of one topic. \\
Version Listing \& Inquiry & Latest, oldest, or count of available versions of a project; drawn from the changelog bucket. \\
Change Retrieval (explicit) & Grounded in a single changelog page; asks about an explicitly listed change. \\
Change Retrieval (implicit) & Asks in which version a behavioral change first appeared; grounded in a chain of up to six consecutive version diffs. \\
\bottomrule
\end{tabular}%
\label{tab:catdefs}
\end{table*}

\subsection{Gold-Answer Audit}
\label{app:goldaudit}

The validation described in Section~\ref{sec:dataset} runs during generation, so it describes how the items were produced, not how correct the released data is. To estimate the error rate of the
released gold answers we drew a stratified sample of 120 items, 10 per category per domain
(fixed seed), and two annotators judged each gold against the released corpus following a written
codebook with four labels: \emph{correct}, \emph{wrong}, \emph{ambiguous} (the gold is only partly
right, or the question admits more than one defensible answer) and \emph{unverifiable}. Annotators
worked in an interface that lists each lineage's versions in release order, counts a search term in
every version, and diffs consecutive versions; every label records the corpus location it rests on,
and neither annotator saw the other's labels while annotating. The 40 Change Retrieval~(i) and
Version Listing items, the two categories that turn on version order, were labeled by both
annotators; the remaining 80 were split between them, stratified by domain and category. The
second annotator reviewed every single-annotator label other than \emph{correct}, and the one
disagreement was resolved by discussion.

On the 40 doubly labeled items the annotators agree on 39 verdicts (97.5\%; Cohen's
$\kappa=0.79$, Gwet's AC1 $=0.97$; $\kappa$ is depressed by the skew toward \emph{correct}).
Two of the 120 golds are wrong: an error rate of 1.7\%, with a 95\% Wilson interval of
[0.5, 5.9], decomposing into 3.3\% on VersionQA-Software and 0.0\% on VersionQA-Legal. Both are
implicit-change items whose gold follows a wrong version order: one places a release candidate
after the release of the same number, the other names v1.32 for a line already present in v1.31,
the earliest indexed version. Two further golds, one per domain, are partly correct: one adds a
rationale its source does not state, the other answers a question that names no version for one
version only. Counting these, 3.3\% [1.3, 8.3] of golds are wrong or partly correct. Reported
accuracies use the original golds.

For implicit-change items the annotators also recorded whether some file states the transition in
words. This holds for 1 of 10 sampled software items but for 9 of 10 legal items, where an earlier
point-in-time version carries an amendment annotation such as ``inserted (1.10.2012)''
(Appendix~\ref{app:legalexample}). Legal implicit-change items are therefore implicit with respect
to the version diff but not necessarily with respect to the indexed text. We release the audit
sample, both annotators' labels with their evidence locations, and the codebook alongside the
datasets.

\subsection{Generation Prompt Templates}
\label{app:genprompts}

Each category has one fixed, author-written template; the generator's only freedom is the pair it
produces from the sources rendered into it. All templates share the skeleton below, differing only
in their \textsc{Task} and \textsc{Constraints} blocks (Figures~\ref{fig:promptcontent}
and~\ref{fig:promptchange}) and in the shape of the payload rendered
into \texttt{\{input\_json\}}: one version's text for the two content types, the list of indexed
version dates for version listing, one pairwise diff for explicit change, the full chain of
consecutive transitions for implicit change.

\begin{codeblock}\begin{verbatim}
You are creating evaluation Q-A pairs for a
version-aware legal-RAG system.

# Task         <category-specific, below>
# Constraints  <category-specific, below>
# Few-shot example
Input:  { ...one worked payload of this
          category's shape... }
Output: { "question": "...", "answer": "..." }

# Output format
Return ONLY a JSON object with keys
"question" and "answer". No prose, no
markdown fences, no explanation.

# Input
{input_json}
\end{verbatim}\end{codeblock}

\begin{figure*}[!t]
\begin{codeblock}\begin{verbatim}
CONTENT RETRIEVAL
Task: Given the legal document text below, generate ONE general-content question and ONE
  answer about a substantive provision of the Act. The question MUST be answerable from the
  text alone, must NOT reference any specific version or date, and should test understanding
  of what the law says.
Constraints: The answer must quote or directly paraphrase the provided text -- do NOT invent
  provisions that are not present. The question must be specific enough to have one clear
  correct answer.

CONTENT RETRIEVAL VERSION-SPECIFIC
Task: Given the legal document text from ONE specific point-in-time version of an Act,
  generate ONE question and ONE answer that test version-specific content. The question MUST
  explicitly reference the version date AND ask about content that is present in that
  version. The answer MUST also reference the version date.
Constraints: The question must mention the act title and the version date (e.g. "What did
  Section 7 of X say in the YYYY-MM-DD version?"). The answer must paraphrase or quote text
  from the provided content, and explicitly reference the version date. Do NOT invent content
  that is not present in the text.

VERSION LISTING & INQUIRY
Task: Given the act title and the list of indexed point-in-time versions, generate ONE
  meta-question about which versions of the Act are available, and ONE answer that lists
  those versions. The question must test the system's ability to enumerate versions, not the
  legal content.
Constraints: Question phrasing variants are acceptable: "Which versions of X are indexed?",
  "How many versions of X do you have?", "List all available versions of X." The answer must
  list ALL the dates from the input. Do NOT invent versions that are not in the input list.
\end{verbatim}\end{codeblock}
\caption{\textsc{Task} and \textsc{Constraints} blocks for the three non-change categories
(legal benchmark). The software templates differ only in rendering project and release
identifiers in place of act titles and in-force dates.}
\label{fig:promptcontent}
\end{figure*}

\begin{figure*}[!t]
\begin{codeblock}\begin{verbatim}
CHANGE RETRIEVAL (EXPLICIT)
Task: Given a textual diff between two point-in-time versions of an Act, generate ONE
  question and ONE answer about a documented amendment described in the newer version's
  amendment notes / footnotes / commencement information. The question must explicitly
  reference the act title and at least one of the two version dates, and the answer must
  describe a concrete change that is mentioned in the diff text. This type focuses on
  EXPLICIT changes -- ones that are textually labelled in the document (e.g., "inserted by
  2020 c. 12 s. 5", "Amendment Note", "Commencement").
Constraints: The question must reference at least one specific version date (the "version_to"
  date is preferred since that's where the change is recorded). The answer must describe what
  was added/removed/substituted, citing the section/paragraph if visible. Do NOT invent
  amendments that are not in the diff text. If no explicit amendment marker is found in the
  diff, write a question that asks about an amendment described in a footnote -- not a
  generic content question.

CHANGE RETRIEVAL (IMPLICIT)
Task: Given pairwise textual diffs between **consecutive** point-in-time versions of an Act,
  generate ONE question and ONE answer about an IMPLICIT change -- a change in the legal text
  itself (added/removed provision, modified definition) that is NOT necessarily labelled with
  an amendment-marker footnote. The question must ask "in which version was X
  added/removed/changed". The answer must specify the EXACT version date in which the change
  first becomes observable.
Constraints: Pick a substantive change (>= 1 sentence of new or removed material) that you
  can localize to **exactly one** transition in the `transitions` list -- i.e. it appears in
  the `to` version of that transition but not in the `from` version, and it does not also
  first occur at any other transition. The answer must name the `to` date of that single
  transition. **Never default to the latest version unless the change actually first appears
  at the very last transition.** The question must phrase the change in terms of WHEN
  something was added/removed/altered (do NOT mention "section X was substituted by 2020
  c. 12" -- that style is for the explicit type). Do NOT invent provisions that are not in
  the diffs. Skip trivial changes (whitespace, citation reformatting, footnote-marker
  renumbering, S.I. number-only updates).
\end{verbatim}\end{codeblock}
\caption{\textsc{Task} and \textsc{Constraints} blocks for the two change categories, which are
separated by question form rather than by source.}
\label{fig:promptchange}
\end{figure*}

Three constraints carry weight. Every template forbids producing anything absent from the rendered
sources, so the later re-scoring against those same sources can check every answer. The
implicit-change template requires the change to be localisable to exactly one transition and
explicitly forbids defaulting to the newest version (the failure mode that would make the
category answerable by always naming the latest date), and excludes whitespace, renumbering, and
citation-reformatting diffs. And the two change templates are separated by question \emph{form},
not by source: explicit names a version date and asks what differs, implicit names the change and
asks when it first appears.

\subsection{Question and Answer Examples}

Tables~\ref{tab:swexamples} and~\ref{tab:legalexamples} give one question--gold pair per
category for each dataset, taken verbatim from the released evaluation sets.

\begin{table*}[t]
\centering
\small
\caption{One example per category from VersionQA-Software, verbatim from the release.}
\begin{tabular}{p{0.17\textwidth}p{0.40\textwidth}p{0.35\textwidth}}
\toprule
Category & Question & Gold answer \\
\midrule
Content Retrieval &
What happens when the Add method is called on a WaitGroup with a delta that causes the counter to go negative? &
If the counter goes negative, Add panics. \\
\addlinespace
Content Retrieval Complex &
How do kind and minikube compare as tools for running Kubernetes locally, and what are their key similarities and differences? &
Both kind and minikube allow users to run Kubernetes on a local computer [\dots] The key difference is that kind requires either Docker or Podman to be installed as a prerequisite, while minikube can run [\dots] \\
\addlinespace
Content Retrieval Version-Specific &
What is the stability level of the \texttt{async\_hooks} module in Node.js version 8.x? &
In Node.js v8.x, the \texttt{async\_hooks} module has a stability level of 1 - Experimental. \\
\addlinespace
Version Listing \& Inquiry &
What is the oldest \texttt{java\_jdk} version you know of? &
17 \\
\addlinespace
Change Retrieval (explicit) &
What dependency was upgraded in Docker Compose 2.38.1? &
\texttt{compose-go} was bumped to v2.7.1 in Docker Compose 2.38.1. \\
\addlinespace
Change Retrieval (implicit) &
In which Go version was the \texttt{strings.Lines} function added? &
The \texttt{strings.Lines} function was added in Go version 1.24.0. \\
\bottomrule
\end{tabular}
\label{tab:swexamples}
\end{table*}

Version-specific content retrieval also contains the feature-absence items described above,
whose gold is the string \texttt{No information available.}; 112 of the 143 items in
that category are positive and the remainder are of this form.

\begin{table*}[t]
\centering
\small
\caption{One example per category from VersionQA-Legal, verbatim from the release.}
\begin{tabular}{p{0.17\textwidth}p{0.40\textwidth}p{0.35\textwidth}}
\toprule
Category & Question & Gold answer \\
\midrule
Content Retrieval &
Which protected characteristics are excluded from Part 6 Chapter 1 of the Equality Act 2010? &
Part 6 Chapter 1 does not apply to age and marriage and civil partnership. \\
\addlinespace
Content Retrieval Complex &
How does section 716(4) of the Income Tax (Earnings and Pensions) Act 2003 define ``the employment change'' for the purpose of applying an order altering the limit under section 287(1), and where is that definition to be found? &
Section 716(4) states that ``the employment change'' has the same meaning as in Chapter 7 of Part 4 (see section 275). Therefore, the definition [\dots] is located in section 275 of the Act. \\
\addlinespace
Content Retrieval Version-Specific &
Under section 1 of the Localism Act 2011 as in force on 15 November 2011, what is the general power of competence conferred on a local authority? &
Under section 1(1) of the Localism Act 2011 as in force on 15 November 2011, a local authority has power to do anything that individuals generally may do. \\
\addlinespace
Version Listing \& Inquiry &
How many versions of Localism Act 2011 Part 6 are you aware of? &
There are four versions of Localism Act 2011 Part 6. \\
\addlinespace
Change Retrieval (explicit) &
What change was made to paragraph 11(2C) of Schedule 15 to the Finance Act 2003 between the 2014-07-17 version and the 2025-03-19 version? &
In the 2025-03-19 version, the text ``or Schedule 4A or 6B'' in paragraph 11(2C) was amended to ``or Schedule 4A~\dots'', omitting the reference to ``6B'' that was present in the 2014-07-17 version. \\
\addlinespace
Change Retrieval (implicit) &
In which version of the Income Tax (Earnings and Pensions) Act 2003 Schedule 5 Part 7 was paragraph 48(3) omitted? &
Paragraph 48(3) was omitted in the version dated 2009-04-01. \\
\bottomrule
\end{tabular}
\label{tab:legalexamples}
\end{table*}

Two properties of the legal set are visible in these examples. Its answers are not short
targets: the median gold is 281 characters and 437 of 547 exceed 120, which is why the judge
is instructed to accept minor paraphrase. And the two change categories are distinguished by
question \emph{form} rather than by source: an explicit item names both versions and asks
what differs between them, whereas an implicit item names the change and asks which version
introduced it. Unlike software, where the split follows the changelog/documentation
partition, both legal categories are derived from the same diffs.

\subsection{Why the Legal Task Is Harder}
\label{app:legalhard}

Three corpus properties, referenced from \S\ref{sub:legal}, make legal a harder operating
point than software rather than the same task with worse numbers. Statutory prose is dense
and self-similar: consecutive versions of an act repeat near-identical boilerplate, so the
retrieval signal separating them is far weaker than between two software documentation
pages. There are no changelogs, so every change must be recovered as an implicit diff
between two long texts rather than read off an explicit entry, which is also why both legal
change categories derive from the same diffs. And amendments are typically micro-edits (a
redefined term, an inserted subsection) embedded in documents orders of magnitude longer
than the edit itself, so the diff signal is small relative to the document carrying it. The
worked example below measures all three on one retained version pair.

\subsection{A Worked Legal Amendment}
\label{app:legalexample}

Schedule 4, Part 1 of the Localism Act 2011 is retained in the corpus at four version dates.
Between the consecutive versions of 2012-11-22 and 2022-05-05, the two files are 104.9\,KB
and 106.0\,KB and their enacted text differs in exactly one place, paragraph 38(2):

\begin{quote}\small
\textbf{2012-11-22:} ``(2) In subsection (2)(a), after \emph{63} insert \emph{as those
sections had effect immediately before their repeal by the Localism Act 2011}.''

\textbf{2022-05-05:} ``[F1](2) \dots\dots\dots\dots'', with an accompanying annotation:
``Sch. 4 para. 38(2) omitted (5.5.2022) by virtue of Local Government and Elections (Wales)
Act \dots''
\end{quote}

Four non-empty lines change out of 351. This is the shape of amendment the legal domain
consists of: an edit three orders of magnitude smaller than the document carrying it,
surrounded by boilerplate that is byte-identical across versions.

Two corpus-level qualifications follow from measuring all 259 consecutive version pairs.
First, retained neighbors are not adjacent amendment events, since the four-version cap spaces
them years apart, so a pair typically aggregates many amendments: the median pair changes
21.4\% of its non-empty lines, and only 14 pairs change under 1\%. The single-subsection case
above is the clean end of a wide distribution, not the typical one. Second, 12 of the 259
pairs (4.6\%) differ in \emph{no} enacted text at all, only in editorial ``Valid from''
markers.

\paragraph{Amendment annotations are present in the indexed text.}
As the example shows, legislation.gov.uk records amendments in \textsc{Textual Amendments}
and \textsc{Commencement Information} annotation blocks, and these are retained in the
converted Markdown: 257 of 355 legal files contain at least one, and in 161 of the 259
consecutive pairs (62.2\%) the newer version's own added text states an amendment dated
exactly that version's date (``S.~7(1)(ba) inserted (25.1.2023) by \dots''). The legal documents therefore do contain a per-section record of their changes, so version-scoped vector search alone can partly answer legal change questions.

\end{document}